\documentclass[letterpaper]{article}
\usepackage{jheppub_mod}
\pdfoutput=1
\usepackage{graphicx, caption, subcaption}  
\usepackage{dcolumn}   
\usepackage{amsmath, amssymb}       
\usepackage{color,latexsym, array,multirow,  verbatim, enumerate}
\usepackage{url}
\usepackage{ulem}
\usepackage{hyperref}

\newcommand{\MuMess}{\Lambda_{\text{mess}}}
\newcommand{\MuInt}{\Lambda_{\text{int}}}
\newcommand{\MuIR}{\mu_{\text{IR}}}
\newcommand{\VecOp}{\mathcal{R}}
\newcommand{\Dtm}{\mathcal{D} } 
\newcommand{\gev}{~\text{GeV} }
\newcommand{\tev}{~\text{TeV} }

\title{Charting generalized supersoft supersymmetry}
\author[a]{Sabyasachi Chakraborty,} 
\author[b]{Adam Martin,}
\author[a,c]{and Tuhin S.~Roy}
\affiliation[a]{Department of Theoretical Physics, Tata Institute of 
Fundamental Research, Mumbai 400005, India}
\affiliation[b]{Department of Physics, University of Notre Dame, IN 46556, USA}
\affiliation[c]{Theory Division T-2, Los Alamos National laboratory, Los Alamos, NM 87545, USA}

\emailAdd{sabya@theory.tifr.res.in, amarti41@nd.edu, tuhin@theory.tifr.res.in}
\date{\today}
\abstract{
Without any shred of evidence for new physics from LHC,  the last hiding spots of natural electroweak supersymmetry seem to lie either in compressed spectra or in spectra where scalars are suppressed with respect to the gauginos. While in the MSSM (or in any theory where supersymmetry is broken by the $F$-vev of a chiral spurion), a hierarchy between scalar and gaugino masses requires special constructions, it is automatic in scenarios where supersymmetry is broken by $D$-vev of a real spurion. In the latter framework,  gaugino mediated contributions to scalar soft masses are finite (loop suppressed but not $\log$-enhanced), a feature often referred to as ``supersoftness". Though phenomenologically attractive, pure supersoft models suffer from the $\mu$-problem, potential color-breaking minima, large $T$-parameter, etc.  These problems can be overcome without sacrificing the model's virtues by departing from pure supersoftness and including $\mu$-type operators that use the same $D$-vev, a framework known as generalized supersoft supersymmetry. The main purpose of this paper is to point out that the new operators also solve the last remaining issue associated with supersoft spectra, namely that a right handed (RH) slepton is predicted to be  the lightest superpartner, rendering the setup cosmologically unfeasible. In particular, we show that the $\mu$-operators in generalized supersoft generate a new source for scalar masses, which can raise the RH-slepton mass above bino due to corrections from renormalisation group evolutions (RGEs). In fact, a mild tuning can open up the bino--RH slepton coannihilation regime for a thermal dark matter. We derive the full set of RGEs required to determine the spectrum at low energies. Beginning with input conditions at a high scale, we show that completely viable spectra can be achieved.  
 }

\begin{document}
\maketitle
\newpage
\section{\label{sec:1}Introduction}
Supersymmetry at the electroweak scale remains one of the most celebrated solution to the hierarchy problem of the Standard Model (SM) to date. However, the LHC experiment have not yet found any excesses in their pursuit of superpartners. After the most recent run, containing an integrated luminosity of nearly 36 fb$^{-1}$ at a center of mass energy of 13 TeV, the constraints on the superpartner masses are becoming quite stringent. For example, the lack of events in the jets plus missing energy search has excluded degenerate squark/gluino scenarios up to masses of 2 TeV, and squarks (including the stop) are now ruled out up to a TeV or so provided they can decay to comparatively lighter neutralinos~\cite{Aaboud:2017iio,Sirunyan:2017cwe,Khachatryan:2016sfv}. However, it should be emphasized that these experimental exclusions are drawn assuming simplified scenarios within the paradigm of the Minimal Supersymmetric Standard Model (MSSM), and are subject to change with more involved topologies.  Within the MSSM, the increased top squark and gluino masses have profound implications for naturalness, as both scales feed into the soft mass of the Higgs via renormalization, dragging it upwards and requiring fine-tuned cancellations among parameters\footnote{Exceptions exist within MSSM where scalar masses can somewhat decouple from gaugino masses. Examples include models with double protection~\cite{Birkedal:2004xi,Chankowski:2004mq,Berezhiani:2005pb,Roy:2005hg,Csaki:2005fc,Bellazzini:2008zy,Bellazzini:2009ix}, scalar sequestering~\cite{Roy:2007nz, Murayama:2007ge, Perez:2008ng}, twin SUSY~\cite{Falkowski:2006qq} etc.}. In short, the promise of a natural MSSM explanation for the weak scale is fading, even if we neglect other generic MSSM problems such as rapid proton decay from dimension five operators, excessive flavor changing neutral currents, and additional CP violating phases.

While the MSSM is the most well studied framework of weak scale supersymmetry, it does not provide all aspects of a general model of electroweak scale supersymmetry. Specifically, the MSSM is an infrared effective supersymmetric framework with superpartner masses sourced by the supersymmetry-breaking vacuum expectation value (vev) of the $F$-component of a chiral spurion. Alternative theories, where supersymmetry breaking is sourced from $D$-vev of a real spurion, are qualitatively distinct and bring several new niceties:
\begin{itemize}
\item 
The primary operators sourced by the $D$-vev generate Dirac masses for gauginos. This requires one to extend the theory to include new chiral superfields in the adjoint representations, as the $D$-vev mass terms pair up the gauginos with the fermion components of the appropriate adjoint superfield, e.g. gluino with color octet fermion, wino with $SU(2)$ triplet fermion, etc. In the absence of any other mass terms, gauginos are purely Dirac. This can be contrasted to the MSSM, where gauginos are purely Majorana.  
\item 
It turns out that one can not write similar $D$-vev sourced operators that can generate scalar masses at the messenger scale.  Further, because of the specific structure of gaugino masses, the gaugino-mediated contribution to scaler masses do not get any $\log$-enhancement and remain finite. This property that the gaugino generated sfermion masses are insensitive to even logarithmic dependence of the ultraviolet (UV) scale, is often referred to as `supersoft'~\cite{Fox:2002bu} -- as opposed to the MSSM, where a logarithmic dependence remains (`soft').  As a result, the scalar masses at the weak scale are loop suppressed with respect to the gaugino masses. 

\item 
The resultant mild split in the squark-gluinos masses are of utmost importance, especially when one calculates the bound in the jets + MET channel.  Heavier gluinos imply suppressed gluino pair and squark-gluino pair production.  Additionally, pair production of same-chirality squarks ($pp \to \tilde q_L \tilde q_L$) is not possible if gluinos are pure Dirac, as it requires a chirality flipping Majorana mass insertion. The reduced cross-sections relax the bound on the first two generation squark masses~\cite{Kribs:2012gx,diCortona:2016fsn}.
\item 
As the gaugino are Dirac particles, the theory (minus the Higgs sector) respects a continuous $U(1)_R$ symmetry. This symmetry suppresses supersymmetric contributions to flavor changing neutral currents and electric dipole moments~\cite{Fox:2002bu, Nelson:2002ca}. More model building can elevate the $U(1)_R$ symmetry to be a symmetry of the full Lagrangian, which ameliorates many of the flavor and CP difficulties of the prototypical MSSM~\cite{Kribs:2007ac, Kribs:2010md} .
\end{itemize}

While the above features are attractive, the supersoft framework does have its share of issues:
\begin{itemize}
\item
The `$\mu$' problem is severe for supersoft models. A non-zero $\mu$ term is essential in a supersymmetry framework as it controls the mass of the charginos. The LEP experiments have placed (nearly) model independent bounds on light charged states, e.g., $m_{\tilde\chi_1^+}>97$ GeV at 95\% C.L~\cite{LEP:2}, therefore a small $\mu$ term is catastrophic. The most elegant solution -- at least in the framework of the MSSM -- is given by the Guidice-Masiero mechanism~\cite{Giudice:1988yz}, however this mechanism requires an $F$-type spurion and thus will not work for supersoft models. Another solution was proposed in Ref.~\cite{Fox:2002bu}, where the conformal compensator field generates the mass for the Higgsinos. While viable, this mechanism relies on a conspiracy among the supersymmetry breaking scale and the Planck scale. A third solution is to include a gauge singlet $S$ in the theory and the superpotential interaction $S H_u H_d$ (e.g., NMSSM~\cite{Ellwanger:2009dp}) which becomes an effective $\mu$-term once the scalar component of $S$ acquires a vev. Supersoft models automatically include the gauge singlet required for this approach as the Dirac partner of the bino. Gauge singlets do require care, however, and may lead to power law UV sensitivity~\cite{Bagger:1995ay}. 
\item 
The infamous `lemon-twist' operators~\cite{Fox:2002bu,Arvanitaki:2013yja,Csaki:2013fla} can break color or generate a too large a $T$-parameter.  Upon replacing the real spurion by its $D$-vev, one finds mass squared of the adjoint scalars to be linear in the coupling constants of these operator. In a large part of the parameter space, the imaginary components of these adjoints may acquire vevs leading to dangerous charge and color breaking vacuum.  Previous solutions to this problem 
require deviation from the supersoftness, resulting in a low energy theory with Majorana gauginos
or extended messenger sectors~\cite{Csaki:2013fla,Amigo:2008rc}. A viable solution of this
problem was put forward in terms of Goldstone gauginos~\cite{Alves:2015kia}, where the right 
handed gaugino is a pseudo-Goldstone field originating from the spontaneous breaking of an 
anomalous flavor symmetry. 
\item
In the framework of pure supersoft supersymmetry breaking, $D$-flatness is a natural direction, leading to vanishing $D$ terms in the potential~\cite{Fox:2002bu}. In the generic MSSM, the $D$-term contribution leads to a tree level mass of the Higgs boson
as $M_Z \cos 2\beta$. Given the discovery of a Higgs with mass close to 125 GeV, a vanishing $D$-term is not a good starting point, as one needs to depend on large quartic corrections at the tree level (NMSSM like) or at the one-loop level to achieve the correct value. Most  $R$-symmetric models with Dirac gauginos therefore include both $F$- and $D$-type breaking to enhance the Higgs quartic term, which consequently increases the Higgs mass. A partial list of such frameworks can be found in the literature~\cite{Fayet:1976et,Polchinski:1982an,Hall:1990dga,Hall:1990hq,Chacko:2004mi,Antoniadis:2005em,Antoniadis:2006eb,Benakli:2008pg,Blechman:2009if,Benakli:2009mk,Kumar:2009sf,Dobrescu:2010mk,Benakli:2010gi,Fok:2010vk,Choi:2010gc,Carpenter:2010as,Abel:2011dc,Benakli:2011kz,Kalinowski:2011zzc,Benakli:2011vb,Frugiuele:2011mh,Brust:2011tb,Davies:2011js,Heikinheimo:2011fk,Bertuzzo:2012su,Davies:2012vu,Goodsell:2012fm,Frugiuele:2012pe, Frugiuele:2012kp,Benakli:2012cy,Itoyama:2013sn,Chakraborty:2013gea,Dudas:2013gga,Itoyama:2013vxa,Beauchesne:2014pra,Benakli:2014cia,Chakraborty:2014tma,Goodsell:2014dia,Ipek:2014moa,Diessner:2014ksa,Carpenter:2015mna,Diessner:2015yna,Goodsell:2015ura,Martin:2015eca,Chakraborty:2015wga,Ding:2015epa,Biggio:2016sdu,Chakraborty:2017dfg,Beauchesne:2017jou,Carpenter:2017xru}.
\item
Another problematic issue of the supersoft scenario is Dark matter (DM). If the $D$-vev spurion remains the only source of superpartner masses and the mediation scale is high, then the lightest supersymmetric particle (LSP) is a right handed (RH) slepton, leading to contradictions with cosmological observations. Lowering the mediation scale, the gravitino becomes the LSP, but this brings its own issues: i.) a lower mediation scale means the conformal compensator mechanism for generating a $\mu$ fails, and ii.) the gravitino LSP scenario is highly constrained because of the long lived charged particle searches~\cite{Kats:2011qh} or  -- if the gravitino mass is less than a few keV -- constraints arise from prompt multi-jet or multi-lepton final states~\cite{ATLAS:2017grs,ATLAS:2017orb}.  
\end{itemize}

A recently proposed framework, `generalized supersoft supersymmetry' (GSS)~\cite{Nelson:2015cea} is free of many of these issues yet maintains key phenomenological features of pure supersoft models, such as finite gaugino mediated contribution to scalar soft masses. Crucially, GSS ameliorates the above issues without introducing additional matter or additional sources of supersymmetry breaking. Instead, in GSS one adds a set of new operators formed from matter superfields and the same $D$-type spurion used to give gauginos mass. When written with chiral adjoints, the new operators give supersymmetric masses to the adjoints and allow one to avoid color-breaking and $T$-parameter issues. When written using Higgses, one finds two $\mu$-parameters (namely, $\mu_u$ and $\mu_d$) are generated, both of which scale with the supersymmetry breaking $D$-vev. If $\mu_u = \mu_d$, these operators can be replaced by the supersymmetric $\mu$-term with $\mu = \mu_u = \mu_d$, which solves the $\mu$ problem in the same spirit as that of Guidice-Masiero mechanism. However when $\mu_u \neq \mu_d$, supersoftness is lost as the hyperchage $D$-term gets a vev and $\log$ sensitivity (\textit{i.e.}, soft) creeps back into the scalar masses. 

The goal of this paper is to flesh out the spectrum of generalized supersoft theories and to determine the parameter regions where they can be viable. While at first glance the loss of supersoftness when $\mu_u \neq \mu_d$ appears to be a flaw in the idea, we will show that it turns out to be a feature that can be utilized to resolve the DM issue that plagues pure supersoft setups. It turns out that this theory with $\Delta\equiv\mu_u-\mu_d\neq 0$ is equivalent to the usual supersoft picture with a supersymmetric `$\mu$' term $\left(\mu=\frac{\mu_u+\mu_d}{2}\right)$, a hypercharge $D$-term with $D_Y\propto \mu\Delta$, and non-holomorphic trilinears (such as $\Delta h_u^{\dagger}\tilde q\tilde u$, etc.). In particular the non-zero $D$-terms provide a
boost to all scalar soft masses, while $D$-term operators involving the adjoint superfields allow more flexibility in the gaugino masses. These effects combine to open up a swath of parameter space where the theory satisfies the observed Higgs mass and achieves the correct thermal relic Dark Matter abundance through bino-slepton coannihilation, all in addition to the usual supersoft phenomenological benefits. Additionally, whenever $\mu_u \neq \mu_d$, non-standard supersymmetry breaking operators arise which give subtle effects in the running of soft parameters. To map out the effects of the GSS UV inputs, we derive the full set of renormalization group equations (RGEs) and provide full numerical results for important outputs such as the Higgs mass and relic abundance. It turns out that RGEs with these unconventional operators are quite subtle and previous attempts~\cite{Jack:1999ud,Jack:2004dv} miss several important effects. To give a more intuitive understandings of electroweak parameters in terms of high scale inputs, we also derive analytical solutions to the RGE under simplifying assumptions.   

The rest of this paper is organized as follows: in Section \ref{sec:2} we begin by laying out the fields, scales, and operators we will consider. Next, in Section \ref{sec:3} we examine how the radiative generation of soft masses in our theory differs from the conventional MSSM and pure supersoft setups, using a toy model to illustrate the key features. In Section~\ref{sec:4} and \ref{sec:5}, we turn to the IR spectra. Many of the gross features of the spectrum can be understood using simplified renormalization group equations that admit analytic solutions, however we will resort to numerics when calculation quantities like the Higgs mass and Dark Matter constraints. This section concludes with a few benchmark points that pass all tests. Finally, a summary and discussion of our results is given in Section~\ref{sec:6} .
\section{\label{sec:2}The Framework:}

The skeleton of any weak scale model of supersymmetry typically consists of two sectors---one containing the MSSM fields, and the other, known as the hidden sector, responsible for supersymmetry breaking, linked by a ``messenger'' sector\footnote{Examples of single site model do exist, see Ref.~\cite{Luty:1998vr}.}.  The messenger sector provides a scale (namely, $\MuMess$) that characterizes all contact operators among the hidden and visible fields. The supersymmetry breaking scale (or rather the supersymmetry breaking vev, namely $\MuInt$) is generated in the hidden sector. Note that the observables at the electroweak scale are various superpartner masses, which are functions of both $\MuInt$ and $\MuMess$, as well as dimensionless numbers, that represent details of messenger mechanisms, and renormalization effects. In fact, renormalization can be quite tricky especially  if the $\MuMess$ is taken as the input scale. As shown in Ref.~\cite{Cohen:2006qc,Roy:2007nz}, the superpartner masses are renormalized  due to interactions of hidden sectors from the scale $\MuMess$ to $\MuInt$, in addition to the usual effects from visible sector interactions. The lack of knowledge of the exact nature of hidden sector dynamics can actually  lead to order one uncertainty in the superpartner masses as calculated using mass-market softwares such as~\texttt{SOFTSUSY}~\cite{Allanach:2001kg}, \texttt{SPheno}~\cite{Porod:2003um,Porod:2011nf}, \texttt{SuSpect}~\cite{Djouadi:2002ze} etc, which completely neglect these effects. Therefore, out the two natural choices for setting the UV input scale of the theory ($\MuMess$ or $\MuInt$)  we choose  $\MuInt$;  below this scale the hidden sector decouples and  the superpartner masses renormalize  only due to visible sector interactions. In other words, we use the superpartner masses and couplings at the scale of renormalization $\mu_r = \MuInt$ to be the input conditions in order to evaluate the spectrum at the electroweak scale. Directly using  $\MuInt$  as the input scale naturally raises questions regarding the nature of hidden sector interactions and the details of the messengers that result in the used boundary conditions. We leave this discussion (and further UV completion) for future work.  

We reiterate that $\MuInt$ is the scale at which contact operators are turned into masses for visible sector fields. In order to simplify, we begin with a hidden sector where the supersymmetry breaking is captured in the vev of a single real field  $\VecOp$, with the following conditions:  
\begin{equation}
\VecOp^\dag \ = \ \VecOp \qquad \qquad \text{and} \qquad \qquad \Dtm \ \equiv  \  \bigg\langle \frac{1}{8} D^2  \bar{D}^2 \VecOp  \bigg\rangle \ > 0 \;
\end{equation}
Here, $D$'s are the usual chiral covariant derivative and $\mathcal{D}$ is the gauge auxiliary fields. Further, we redefine $\VecOp \rightarrow \VecOp/\MuMess^d$, where $d$ is the engineering dimension of the operator at $\MuMess$.  This redefinition sets the engineering dimension of $\VecOp$ to be $0$.   

The visible sector in generalized supersoft supersymmetry extended beyond the MSSM content to include three extra chiral superfields in the adjoint representation of the three SM gauge groups, i.e. we include a color octet $\Sigma_3$, a weak triplet $\Sigma_2$, and a charge-neutral $\Sigma_1$.  Given this field content, we will work with the minimal set of contact operators (between the hidden and visible sector fields) capable of generating a viable $\MuIR$ spectrum. First, consider the conventional supersoft  operator~\cite{Fox:2002bu} that gives rise to Dirac type gaugino masses. 
\begin{equation} 
\begin{gathered}
 -   \frac{1}{2}  \int \!\! d^2 \theta \   \omega_i(\MuInt) \frac{1}{\MuMess } \bar{D}^2  D^\alpha \VecOp  \ W_{\alpha, i} \Sigma_i 
 \quad \rightarrow \quad 
 M_{D_i}(\MuInt) \ \lambda_i \psi_i \;, \\
\text{where} \qquad  M_{D_i}(\MuInt) \ = \    \omega_i( \MuInt) \frac{\Dtm}{\MuMess} \,,
\end{gathered}
\label{eq:dirac_gaugino}
\end{equation}
where $W^\alpha_i$ is the field-strength chiral superfields for the $i$-th SM gauge group ($i$ is not summed over, but spinor Lorentz indices $\alpha$ are summed) and $\omega_i$ are dimensionless coupling constants.  
In Eq.~\eqref{eq:dirac_gaugino} we have suppressed any gauge group indices. By design, the supersymmetry breaking vev of $\VecOp$ picks out gauginos ($\lambda_i$) from $W_i$ and the fermionic component of $\Sigma_i$ fields (denoted by  $\psi_i$).   
Another operator often used in this context is known as the `lemon-twist' operators~\cite{Fox:2002bu}: 
\begin{equation}
\begin{gathered}
 -   \frac{1}{2}  \int \!\! d^2 \theta  \ \omega_{q q'} \  \frac{1}{\MuMess^2} \left( \bar{D}^2 D \VecOp \right)^2  \ Q  Q' 
 \quad \rightarrow \quad 
 b_{qq'}(\MuInt) \ \phi_{q} \phi_{q}'  \;, \\ 
\text{where} \qquad  b_{qq'}(\MuInt) \ = \    \omega_{qq'}( \MuInt) \frac{\Dtm^2}{\MuMess^2} \,  , 
\end{gathered}
\label{eq:lemon_twist}
\end{equation}
where $Q$ and $Q'$ represent visible sector chiral superfields with scalar components $\phi_q, \phi'_q$, and the coupling constant $\omega_{qq'}$ is nonzero only if $Q Q'$ is a gauge singlet. Examples of such operators include $\Sigma_i^2$, and $H_u H_d$. These operators are problematic because they give opposite sign mass to the real and imaginary parts of the scalars and thus can drive fields tachyonic~\cite{Fox:2002bu}.

In addition to Eq. \eqref{eq:dirac_gaugino} and \eqref{eq:lemon_twist}, we use the operators proposed in Ref.~\cite{Nelson:2015cea} to generate the $\mu$ term.
\begin{equation} 
\begin{gathered}
 -   \frac{1}{4}  \int \!\! d^2 \theta  \ \omega_{Dq q'} \  \frac{1}{\MuMess} \bar{D}^2 \left( D^\alpha \VecOp  D_\alpha Q\right)  \   Q' 
 \quad \rightarrow \quad 
 \mu_{Dqq'}(\MuInt) \  \big( \frac{1}{2} \psi_{q} \psi_{q'}  \ + \    F_Q \phi_{Q'} \big) \;, \\ 
\text{where} \qquad  \mu_{Dqq'}(\MuInt) \ = \    \omega_{Dqq'}( \MuInt) \frac{\Dtm}{\MuMess} \, ,
\end{gathered}
\label{eq:non_stand}
\end{equation}
where $F_Q$ represents the auxiliary component of the field $Q$.  As in Eq.~\eqref{eq:lemon_twist},  the coupling constant $\omega_{Dqq'}$ are nonzero only if the chiral operator $Q Q'$ is a gauge singlet.  The potential generated after eliminating the auxiliary fields is given by 
\begin{equation}
\Big| \frac{\partial W}{\partial Q }  \ + \      \mu_{Dqq'}(\MuInt)   \phi_{Q'} \Big|^2 \;;
\end{equation}
crucially, if $Q$ and $Q'$ represent different fields, the equation above only gives mass for $Q'$-scalars. For non-zero $\frac{\partial W}{\partial Q } $, Eq.~\eqref{eq:non_stand} also gives rise to trilinear scalar mass terms. As an explicit example, substituting   $Q \rightarrow H_u$ and $Q'\rightarrow H_d$,  Eq.~\eqref{eq:non_stand} generates masses for Higgsinos, the scalar $h_d$, and a non-holomorphic scalar trilinear operator $h_d^\dag \tilde{q} \tilde{u}$.  If we flip $H_u$ and $H_d$ (namely, $Q \rightarrow H_d$ and $Q'\rightarrow H_u$), Eq.~\eqref{eq:non_stand} instead gives rise to masses for the Higgsinos and $h_u$ scalar, and  operator $h_u^\dag \tilde{q} \tilde{d}$. Including both possibilities and denoting the mass scales of the two operators by $ \mu_{d}$ and $ \mu_{u}$, we find the following terms in the Lagrangian:
\begin{equation} 
\mathcal{L} \ \supset \  \frac{1}{2} \left(\mu_d + \mu_d \right)   \: \tilde{H}_u \tilde{H}_d  \ - \  m_{h_u}^2 \ \left|  h_u \right|^2  \ - \  m_{h_d}^2 \ \left|  h_d \right|^2    
\ - \  Y_u \mu_d^* \ h_d^\dag   \tilde{q} \tilde{u} 
\ - \  Y_d \mu_u^* \ h_u^\dag   \tilde{q} \tilde{d}  \ + \text{h.c.}  \, . 
\label{eq:LHiggsOri}
\end{equation}
In the limit $\mu_u = \mu_d = \mu/2$, all the Higgs scalars and the Higgsinos have identical masses, a result identical to what we would get from superpotential $\mu H_u H_d$. To make this supersymmetric limit more apparent we can rewrite Eq.~\eqref{eq:LHiggsOri} in terms of a new superpotential term (namely, the $\mu$-term) and soft supersymmetry breaking mass terms: 
\begin{equation} 
\begin{split}
W \ &\supset \ \mu \: H_u H_d \\
\mathcal{L}_{\text{soft}} \ & \supset \  
- \ \tilde{m}_{h_u}^2 \ \left|  h_u \right|^2  \ - \  \tilde{m}_{h_d}^2 \ \left|  h_d \right|^2    
\ - \  Y_u \tilde{\xi}_u^* \ h_d^\dag   \tilde{q} \tilde{u} 
\ - \  Y_d \tilde{\xi}_d^* \ h_u^\dag   \tilde{q} \tilde{d}  \ + \text{h.c.}  \, , 
\end{split}
\label{eq:LHiggsSoft}
\end{equation}
where $\tilde{m}_{h_u}^2 = \left| \mu_u^0 \right|^2 - \left| \mu^0 \right|^2 $, $\tilde{m}_{h_d}^2 = \left| \mu_d^0 \right|^2 - \left| \mu^0 \right|^2$, and $\tilde{\xi}_u = - \tilde{\xi}_d =   \frac{1}{2} \left( \mu_d^0 - \mu_u^0 \right)$. Written this way,  
the supersymmetric limit corresponds to $\mathcal{L}_{\text{soft}}  \rightarrow 0$. In addition, supersoft models are well motivated as the flavor and CP violating effects are suppressed. In our scenario, we are generating non-standard trilinear scalar terms proportional to $Y_u$ and $Y_d$. As a result, minimal flavor violation (MFV) would ensure that flavor issues are under control.

As pointed out in~\cite{Nelson:2015cea}, the operators  in eq.~\eqref{eq:non_stand} can also be used to solve potential color breaking or a large $T$-parameter issues in supersoft models by preventing scalars from $\Sigma$ fields from becoming tachyonic. In detail, one needs to include the operators in Eq.~\eqref{eq:non_stand}, with $QQ'$ replaced by $\text{Tr}(\Sigma_a^2)$. Repeating the same steps as shown after Eq.~\eqref{eq:non_stand}, this results in a mass for both the fermion and scalar components of $\Sigma_a$. The Majorana mass for $\psi_a$ upsets the cancellation of the $SU(2)$ and $U(1)$ D-terms that occurs in pure supersoft scenarios, regenerating the tree-level Higgs quartic. Additionally, the scalar mass squared from Eq.~\eqref{eq:non_stand} is positive for {\it both} the real and imaginary parts of the adjoint scalar, thus the issue of tachyonic masses can be avoided provided contributions from Eq.~\eqref{eq:non_stand} are larger than any `lemon twist' adjoint masses.

At this point, let us reiterate that we are still describing the theory at the scale of renormalization $\mu_r = \MuIR$. In order to calculate the spectrum at the electroweak scale, these operators must be renormalized below $\MuInt$. Renormalization gives rise to new counter-terms and changes the couplings of various operators. In particular, one finds that the nice relationship between $\mu, \tilde{\xi}_{u,d}, \tilde{m}_{h_{u,d}}$, are modified and need to be treated independently.  We therefore, collect all the terms described above, include new operators to account for generated counter-terms as follows: 
\begin{equation}  
\begin{split}
W &\ = \ Y_u  \: QH_uU  \ + \  Y_d \:  QH_d D \ + \ Y_{\tau} \: LH_d E \ + \ \mu \: H_u H_d  \ + \ \frac{1}{2} \sum_a M_{\Sigma_a} \: \text{Tr}\left( \Sigma_a^2 \right) \\
\mathcal{L}_{\text{soft}}   &\   \supset \  \sum_a \  M_{D_a} \:  \lambda_a \psi_a \  - \ \sum_{\phi} \: \tilde{m}_{\phi}^2 \ \left| \phi \right|^2 
	\ - \  \tilde{m}_{h_u}^2  \: \left| h_u \right|^2 \ - \  \tilde{m}_{h_d}^2  \: \left| h_d \right|^2         \\
& \qquad \ - \  Y_u \tilde{\xi}_u^* \: h_d^\dag   \tilde{q} \tilde{u}    
\ - \  Y_d  \tilde{\xi}_d^* \: \ h_u^\dag   \tilde{q} \tilde{d} 
\ - \  Y_\tau \tilde{\xi}_\tau^* \: h_u^\dag   \tilde{\ell} \tilde{e}
\ - \  b_\mu \: h_u h_d 
 \ + \ \text{h.c.}  \, ,
\label{eq:GSSLagrangian}
\end{split}
\end{equation}
where  $\phi $ stands for $ \tilde{q}_i, \tilde{u}_i, \tilde{d}_i, \tilde{l}_i, \tilde{e}_i $, and $i$ denotes the family. Coefficients of many of these operators are either zero or related to each other at the input scale $\MuInt$, as discussed before. 

Before we proceed any further, let us completely specify the coefficients at the input scale as well as establish our notation. Our UV inputs are:
\begin{itemize}

\item The Dirac gaugino masses and supersymmetric masses for the adjoints ($\Sigma$ fields), as well as the supersymmetric  $\mu$  parameter:
\begin{equation}
M_{D_a} \left( \MuInt \right) \  = \ M_{D_a}^0, \qquad  M_{\Sigma_a} \left( \MuInt \right) \  = \ M_{\Sigma_a}^0,  \qquad  \mu \left( \MuInt \right) \  = \ \mu^0 \label{eq:mu0}\; ,
\end{equation}
where $a$ runs over the three gauge groups in the SM.  

\item Soft masses for scalars   
\begin{equation}
\tilde{m}_{\phi}^2  \left( \MuInt \right) \  = \  0, \qquad 
\tilde{m}_{h_u}^2  \left( \MuInt \right) \  = \  \left| \mu_u^0 \right|^2 - \left| \mu^0 \right|^2 , \qquad
\tilde{m}_{h_d}^2  \left( \MuInt \right) \  = \  \left| \mu_d^0 \right|^2 - \left| \mu^0 \right|^2 \; .
\label{eq:scalar_bound}
\end{equation}
Another important quantity is the Hypercharge $D$-term defined as $\mathcal{S}  \  \equiv \   \sum_{\phi} \: q_\phi  \tilde{m}_{\phi}^2$, where $\phi$ runs over all particles,  and $q_\phi$ represent corresponding hyperchages. Therefore, we find the boundary value of $\mathcal{S}$ to be 
\begin{equation}
\mathcal{S} \left( \MuInt \right)  \ = \  \left( \left| \mu_u^0 \right|^2 -  \left| \mu_d^0 \right|^2 \right)  \ \equiv \  \mathcal{S}_0 \; .
\end{equation}

\item The initial conditions for the soft trilinear $\tilde{\xi}$ operators are completely specified in terms of the Higgs sector parameters: 
\begin{equation}
\begin{split}
& \tilde{\xi}_u \left( \MuInt \right)  \ = \  \frac{1}{2} \left( \mu_d^0 - \mu_u^0 \right)  
 \  \equiv \   	\tilde{\xi}_0\;,   \\
& \tilde{\xi}_{d/\tau} \left( \MuInt \right)  \ = \  
	\frac{1}{2} \left( \mu_u^0 - \mu_d^0 \right)  
 \  \equiv \   - \tilde{\xi}_0.
\end{split} 
\end{equation}
\end{itemize}

We devote the next section for deriving the renormalization group equations for the coupling constants given in Eq.~\eqref{eq:GSSLagrangian}.

\section{Renormalization Group Equations}
\label{sec:3}

With the exception of the non-holomorphic trilinear terms (namely the $\xi$-operators) in Eq.~\eqref{eq:GSSLagrangian}, RGEs of all other operators are well known and widely used. The effect of the $\xi$-operators, on the other hand, are extremely non-trivial and subtle. Early efforts in deriving these missed several effects~\cite{Jack:1999ud,Jack:2004dv}, and the RGEs of these operators and their effects in RGEs of other operators get more complicated in the presence of various Yukawa terms, gauge couplings, and, in particular, the hypercharge $D$-term.  In order to get things correct, we begin with a simple toy model consisting of: 
\begin{enumerate}
\item A minimal set of degrees of freedoms;
\item Only a single Yukawa coupling;
\item Only global symmetries: this assumption allows us to disregard complexities due to gaugino masses. Towards the end we will gauge one of the global $U(1)$ in the model, and study the effects of  $\xi$-operators, in the presence of $D$-term.   
\end{enumerate}
RGEs of the full model of Eq.~\eqref{eq:GSSLagrangian} is given in Sec.~\ref{subsec:3.2}. Readers interested in seeing the final form can simply jump to  Sec.~\ref{subsec:3.2}.

\subsection{A Toy Model}
\label{subsec:3.1}
In order to understand the non-trivial effects of the operator in Eq.~\eqref{eq:non_stand} on we construct a toy model of the visible sector described by five multiplets (namely, $Q, U, D, H_u$ and $H_d$) charged under the \textbf{global} groups $G_1 \times G_2 \times U(1)_X \times U(1)_R$ as described in Table~\ref{table:toy}, along with a single superpotential term (namely, $Y  \: QH_uU$). The non-generic nature of the holomorphic superpotential allows us to get away with not writing any other marginal interaction.     
\begin{table}[h]
\begin{center}
\begin{tabular}{|c|c|c|c|c|}
\hline
 & $G_1 \equiv SU(3)$ & $G_2 \equiv SU(2)$ &  $U(1)_X$ & $U(1)_R$  \\
 \hline
 $Q$ & $3$ & $2$ & 0 & 0 \\
 $U$ & $\bar{3}$ & 1 & -1 & 1 \\
 $D$ & $\bar{3}$ & 1 & 1 & 1 \\
 $H_u$ & 1 & 2 & 1 & 1 \\
 $H_d$ & 1 & 2 & -1 & 1 \\  
 \hline
\end{tabular}
\end{center}
\caption{Charge assignments of the chiral superfields in the toy model. The global symmetries help us to disregard complexities due to the gaugino masses.}
\label{table:toy}
\end{table}

With this particle content, the only invariant, holomorphic bilinear we can form is $H_u H_d$. Consequently, there can be two operators of the form Eq.~\eqref{eq:non_stand}  (namely, one term with derivative acting on $H_u$, and the second with derivative on $H_d$).  At the input scale $\MuInt$, this give rise to masses for Higgsinos, scalar Higgses as well as non-holomorphic scalar trilinear.   
\begin{equation} 
\mu_r = \MuInt : \quad \mathcal{L} \ \supset \  
\frac{1}{2}  \left( \mu_u + \mu_d \right) \ \tilde{H}_u \tilde{H}_d  \ - \ 
\big| \mu_{u} \big|^2 \ \left|  h_u \right|^2  \ - \ 
\big| \mu_{d} \big|^2 \ \left|  h_d \right|^2   \ - \
Y \mu_d^* \ h_d^\dag   \tilde{q} \tilde{u} \ + \ \text{h.c.}.
\label{eq:input-toy}
\end{equation}
In Eq.~\eqref{eq:input-toy} we have explicitly written down $\mu_r = \MuInt $ in order to indicate the fact that the relationships exhibited amongst masses of Higgs scalars, Higgsinos, as well as the couplings is only valid at the scale $\MuInt$.  In order to renormalize the theory below $\MuInt $, additional counter-terms are needed, and this is where the full symmetry structure in Table~\ref{table:toy} is helpful since it restricts the number of operators we need to consider considerably. Further, the $D$ multiplet does not have any interaction at all, and as a result no new counter-term involving $D$ needs to be written down (another way to see it is the fact that with the current interactions, there is an additional $U(1)$ symmetry under which only $D$ is charged, and this restricts more counter-terms). Including all of the counter-terms we need to take into account (and that are invariant under all global symmetries mentioned above), the Lagrangian in this toy model at any scale $\mu_r \leq \MuInt $,  is given by:
\begin{equation}
\begin{gathered}
W \ = \ Y  \: QH_uU \; , \\ 
\mathcal{L} \ \supset \  \mu   \ \tilde{H}_u \tilde{H}_d  \ - \  m_{h_u}^2 \ \left|  h_u \right|^2 
\ - \  m_{h_d}^2 \ \left|  h_d \right|^2  \ - \ m_{\tilde{q}}^2  \left|  \tilde{q} \right|^2 
\ - \ m_{\tilde{u}}^2  \left|  \tilde{u} \right|^2 \ - \ m_{\tilde{d}}^2  \left|  \tilde{d} \right|^2   \\ 
\ - \  Y \xi_u^* \ h_d^\dag   \tilde{q} \tilde{u}  \ + \text{h.c.}  \, . 
\end{gathered}
\label{eq:ToyLagrangian}
\end{equation}
Terms corresponding to traditional $a$-terms, such as $h_u \tilde{q} \tilde{u}$ ($h_u^{\dagger} \tilde{q} \tilde{u}$), or the $b_\mu$-term break $U(1)_R$ ($U(1)_X$) and therefore will not be generated via loops.   Note that the masses-squared parameters in Eq.~\eqref{eq:ToyLagrangian}, such as $m_{h_u}^2$, refer to the full masses of the scalar fields. Following the logic of Eq.~\eqref{eq:LHiggsSoft}, this Lagrangian in can be re-expressed as a superpotential piece plus soft terms:
\begin{equation} 
\begin{gathered}
W \ = \ Y  \: QH_uU \ + \ \mu\:  H_u H_d \; ,\\
\mathcal{L}_{\text{soft}}   \   \supset \ - \  \tilde{m}_{h_u}^2 \ \left|  h_u \right|^2 
\ - \  \tilde{m}_{h_d}^2 \ \left|  h_d \right|^2  \ - \ \tilde{m}_{\tilde{q}}^2  \left|  \tilde{q} \right|^2 
\ - \ \tilde{m}_{\tilde{u}}^2  \left|  \tilde{u} \right|^2 \ - \ \tilde{m}_{\tilde{d}}^2  \left|  \tilde{d} \right|^2   \\ 
\ - \  Y \tilde{\xi}_u^* \ h_d^\dag   \tilde{q} \tilde{u} \ + \text{h.c.} \; , 
\end{gathered}
\label{eq:ToyLagrangianV2}
\end{equation}
with
\begin{equation}
\tilde{m}_{h_u}^2 \ = \ m_{h_u}^2 - \left| \mu \right|^2 \; , \quad 
\tilde{m}_{h_d}^2 \ = \ m_{h_d}^2 - \left| \mu \right|^2 \; , \quad
\tilde{\xi}_u \ = \ \xi_u - \mu  \; , \quad \text{and} \quad
\tilde{m}_{\tilde{q}, \tilde{u}, \tilde{d}}^2 \ = \ m_{\tilde{q}, \tilde{u}, \tilde{d}}^2 \; .
\label{eq:ToySubstitution}
\end{equation}
The supersymmetric limit of Eq.~\eqref{eq:ToyLagrangianV2} is obvious, namely $\mathcal{L}_{\text{soft}}  \rightarrow 0$, while for ~\eqref{eq:ToyLagrangian}, the supersymmetric  limit corresponds to: $m_{h_{u,d}}^2 \rightarrow \left| \mu \right|^2,\,\xi \rightarrow \mu,\, m_{\tilde{q}, \tilde{u}, \tilde{d}}^2  \rightarrow 0$.

It is instructive to derive the RGEs both in term of mass-parameters from Eq.~\eqref{eq:ToyLagrangian} and in terms of soft-mass parameters in Eq.~\eqref{eq:ToyLagrangianV2}, then match the two apprroaches using the substitutions stated above for a consistency check. However, to save space in this write-up we show only one derivation, the $\beta$-functions of the operators in  Eq.~\eqref{eq:ToyLagrangian}; the $\beta$-functions for the soft parameters  can be derived using the the substitutions in Eq.~\eqref{eq:ToySubstitution}

\begin{figure}[h]
\centering
\includegraphics[width=0.85\textwidth]{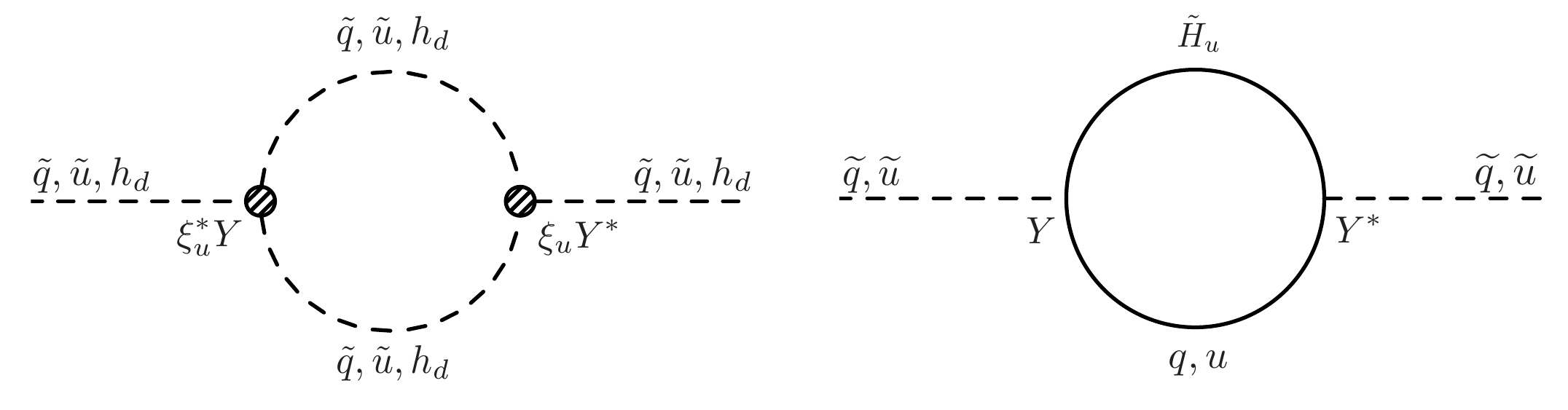}
\caption{Non standard soft supersymmetry breaking terms contributing to the running of the scalar fields.}
\label{fig:first}
\end{figure}
At one loop order, the $\beta$-functions can be evaluated diagrammatically. In addition to the standard diagrams one encounters in MSSM calculations, we need to take into account new diagrams due to the $\xi$ operators (e.g., Fig.~\ref{fig:first}). Starting from Eq.~(\ref{eq:ToyLagrangian}), the full $\beta$-functions for the Yukawa coupling $Y$, the Higgsino mass parameter $\mu$, the non-holomorphic scalar trilinear parameter $\xi_u$, and various scalar mass-squared parameters are given below.   
\begin{align} 
&16\pi^2 \: \beta\left( Y \right) \ = \  6 \left| Y \right|^2 \: Y \; ,  \label{eq:b_y} \\
&16\pi^2 \: \beta\left( \mu \right) \ = \  3 \left| Y \right|^2 \mu   \; , \label{eq:b_mu} \\  
&16\pi^2 \: \beta\left( \xi_u \right) \ =  \ \frac{1}{Y^*} \beta\left( Y^* \xi_u \right) -   \frac{\xi_u}{Y^*} \beta\left( Y^* \right)  \ = \ 3 \left| Y \right|^2 \xi_u   \; , \\
&16\pi^2 \: \beta\left( m_{\tilde{q}}^2 \right)  \ = \ 2 \left| Y \right|^2 \left( m_{\tilde{q}}^2 + m_{\tilde{u}}^2 + m_{h_u}^2 +  \left| \xi_u \right|^2 - 2 \left| \mu \right|^2 \right) \; ,
		\label{eq:b_mq} \\
&16\pi^2 \: \beta\left( m_{\tilde{u}}^2 \right)  \ = \  4 \left| Y \right|^2 \left( m_{\tilde{q}}^2 + m_{\tilde{u}}^2 + m_{h_u}^2 +  \left| \xi_u \right|^2 - 2 \left| \mu \right|^2 \right) \; ,
		\label{eq:b_mmu}  \\
&16\pi^2 \: \beta\left( m_{\tilde{d}}^2 \right)  \ = \ 0 	 \; ,	\label{eq:b_md} \\ 
&16\pi^2 \: \beta\left( m_{h_u}^2 \right)  \ = \  6 \left| Y \right|^2 \left( m_{\tilde{q}}^2 + m_{\tilde{u}}^2 + m_{h_u}^2  \right) 	 \; ,	\label{eq:b_mhu} \\
&16\pi^2 \: \beta\left( m_{h_d}^2 \right)  \ = \  6 \left| Y \right|^2 \left| \xi_u \right|^2  	 \; .	\label{eq:b_mhd} 
\end{align}

\subsubsection{Detour: consistency check}
As a consistency check for the RGEs derived above due to the unconventional $\xi$ operator, consider taking the supersymmetric limit. In particular, we look at the mass parameter $m_{h_d}^2$ -- in the supersymmetric limit,  $m_{h_d}^2$ should be equal to the Higgsino mass parameter (namely,   $\left| \mu \right|^2$) at all scales,
  \begin{equation}
\lim_{\xi \rightarrow \mu} \  \frac{d}{dt} \left(  m_{h_d}^2  -  \left| \mu \right|^2 \right) \ = \ 0 \; . 
\end{equation}
Staring at Eq.~(\ref{eq:b_mu}, \ref{eq:b_mhd}), we see our RGE pass this check.

\subsubsection{Towards the full Lagrangian}
\label{subsec:3.2}
To formulate the RGEs of the full theory as given in Eq.~\eqref{eq:GSSLagrangian}, one needs to incorporate important complexifications:
\begin{itemize}
\item
If the superpotential in Eq.~\eqref{eq:ToyLagrangian} is expanded to include new marginal interactions involving the $D$ superfield, such as $\bar{Y} \: QH_d D$, the accidental global $U(1)$ symmetry with $D$ is lost and new  counter-terms are needed. For example, in the presence of both $Y$ and $\bar Y$, the operator  $ h_d^\dag \tilde{q} \tilde{u}$ shown in Fig.~\ref{fig:sec} is permitted.
\begin{figure}[h]
\centering
\includegraphics[width=0.4\textwidth]{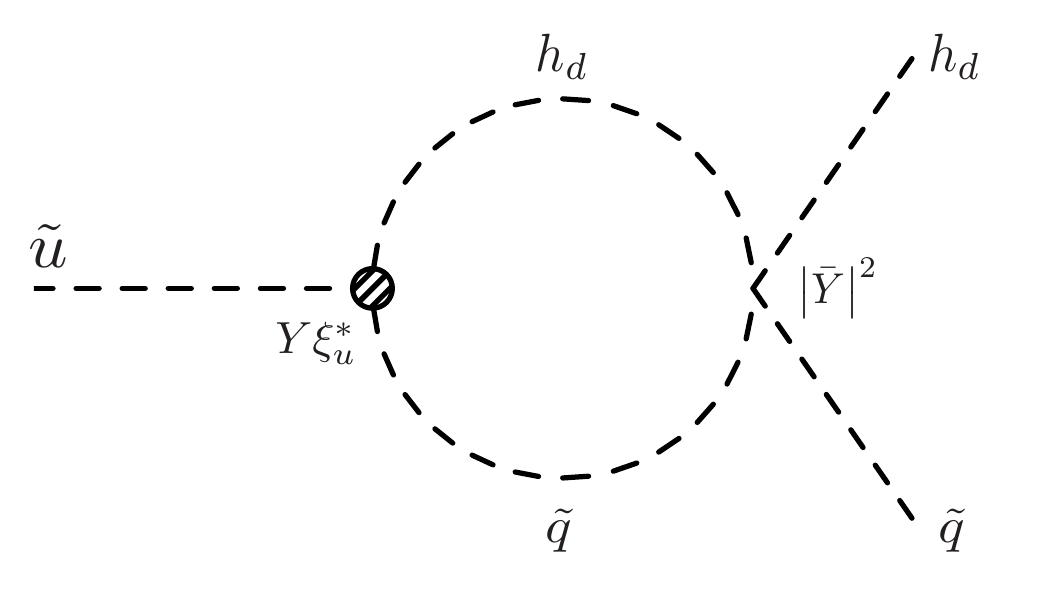}
\caption{Generation of new operator $h_d^{\dagger}\tilde q\tilde u$ because of the presence of $\bar Y$.}
\label{fig:sec}
\end{figure}

\item If one gauges $U(1)_X$, the effect of its $D$-terms need to be taken into account, even if we refrain from adding mass term for the gaugino (so that the $U(1)_R$ remains unbroken, and we can keep using the symmetry arguments in order to restrict counter-terms). The impact of gauging shows up in two places. 
First, the anomalous dimensions of all superfields  charged under $U(1)_X$ changes because of the gauge fields. Second, new additive contributions to the RGEs arise because of the $U(1)_X$  $D$-term. The contribution can be summarized in terms of the parameter $\mathcal{S}_X$.
\begin{align}
&\mathcal{S}_X \ \equiv \  \: \sum_{\phi} q_{\phi} \:  \tilde{m}_{\phi}^2 \;, \\ 
&\beta\left( \tilde{m}_{\phi}^2 \right)  \ \rightarrow  \  \beta\left( \tilde{m}_{\phi}^2 \right)  - g_X^2 \:q_{\phi} \: \mathcal{S}_X \;, \\
&\beta\left( \mathcal{S}_X \right) \ = \  g_X^2 \mathcal{S}_X 
 	\ - \  12  \left| Y \right|^2  \left(  \left| \tilde{\xi}_u \right|^2 + \tilde{\xi}_u \mu^* +  \tilde{\xi}_u^* \mu \right) 
		\ + \  12  \left| \bar{Y} \right|^2 \left(   \left| \tilde{\xi}_d \right|^2  
			+ \tilde{\xi}_d \mu^* +  \tilde{\xi}_d^* \mu \right)   \; ,
\end{align}  
where $\phi$ runs over all scalars of the model, $q_{\phi}$ represents $\phi$'s charge under $U(1)_X$, and $g_X$ represents the gauge coupling constant. 

Importantly, these RGE hold as long as $U(1)_R$ remains unbroken. Therefore, even in the presence of  $U(1)_R$ preserving gaugino mass terms (\textit{i.e.}, Dirac gaugino masses) the above one-loop results prevail.  

\item if one adds a $b_\mu$ term to the toy model, it is multiplicatively renormalized and does not enter the RGE for any other parameters. Both effects stem from the fact that $b_\mu$ is the only $U(1)_R$ breaking term and has the wrong mass dimension to radiatively generate $a$-terms.

\end{itemize}

The experience with the toy model (with or without added complications) paves way for us to write down the RGEs of the full model, as shown in the next section.

\subsection{Renormalization group equations in the full model}
We will take the approximation that only the third generation of the Yukawa couplings are non-zero. This reduces Yukawa coupling matrices to
\begin{equation}
   Y_t \ = \
  \left( {\begin{array}{ccc}
   0 & 0 & 0 \\
   0 & 0 & 0 \\
   0 & 0 & y_t  \\
  \end{array} } \right) \; ,
  \qquad
   Y_b \ = \
  \left( {\begin{array}{ccc}
   0 & 0 & 0 \\
   0 & 0 & 0 \\
   0 & 0 & y_b \\
  \end{array} } \right) \; ,
   \qquad
   Y_\tau \ = \
  \left( {\begin{array}{ccc}
   0 & 0 & 0 \\
   0 & 0 & 0 \\
   0 & 0 & y_\tau \\
  \end{array} } \right) \; .
\end{equation}
When expressing the RGE in the full model, we will work in the more familiar language of running supersymmetric and supersymmetry breaking parameters, in the spirit of Eq.~\eqref{eq:ToyLagrangianV2}. To split RGE for full masses into supersymmetric and soft pieces, one needs to express the soft mass parameters in terms of the mass parameters in Eq.~\eqref{eq:ToyLagrangian} using Eq.~\eqref{eq:ToySubstitution}, then apply equations Eqs.~(\ref{eq:b_y}-\ref{eq:b_mhd}).  Written in this form and using the reduced Yukawa matrices, the $\beta$-functions of the soft masses are: 

\begin{align}
\begin{split}
&16\pi^2 \: \beta\left( \tilde{m}_{\tilde{q}_3}^2 \right)  \ = \ 2 \left| y_t \right|^2 \left( \tilde{m}_{\tilde{q}}^2 + \tilde{m}_{\tilde{u}}^2 + \tilde{m}_{h_u}^2  \right)
		\ + \ 2 \left| y_b \right|^2 \left( \tilde{m}_{\tilde{q}}^2 + \tilde{m}_{\tilde{d}}^2 + \tilde{m}_{h_d}^2  \right) \\ & 
\qquad  \qquad
			\ + \  2 \left| y_t \right|^2 \: \left( \left| \tilde{\xi}_u \right|^2  \ +  \tilde{\xi}_u \mu^* +  \tilde{\xi}_u^* \mu \right)  
					\ + \  2 \left| y_b \right|^2 \:  \left(\left| \tilde{\xi}_d \right|^2 
				\ + \ \tilde{\xi}_d \mu^* +  \tilde{\xi}_d^* \mu \right)\ + \frac{1}{5} g_1^2 \mathcal S  \; , \\ & 
					\label{eq:b2a_tmq} 
\end{split} \\
\begin{split}
 &16\pi^2 \:  \beta\left( \tilde{m}_{\tilde{u}_3}^2 \right)  \ = \  4 \left| y_t \right|^2 \left( \tilde{m}_{\tilde{q}}^2 + \tilde{m}_{\tilde{u}}^2 + \tilde{m}_{h_u}^2  \right)
	\ + \  4 \left| y_t \right|^2 \: \left( \left| \tilde{\xi}_u \right|^2  \ + \   \tilde{\xi}_u \mu^* +  \tilde{\xi}_u^* \mu \right) \ - \ \frac{4}{5}g_1^2 \mathcal S   \; , \\ &
	\label{eq:b2a_tmu} \\
\end{split} \\
\begin{split}
&16\pi^2 \:  \beta\left( \tilde{m}_{\tilde{d}_3}^2 \right)  \ = \ 4  \left| y_b \right|^2 \left( \tilde{m}_{\tilde{q}}^2 + \tilde{m}_{\tilde{d}}^2 + \tilde{m}_{h_d}^2  \right)
	\ + \  4 \left| y_b \right|^2 \:  \left( \left| \tilde{\xi}_d \right|^2 \ + \   \tilde{\xi}_d \mu^* +  \tilde{\xi}_d^* \mu \right) \ + \ \frac{2}{5}g_1^2\mathcal S  \; ,
 \\ &
				\label{eq:b2a_tmd} \\ 
\end{split} \\
\begin{split}
&16\pi^2 \: \beta\left( \tilde{m}_{\tilde{\ell}_3}^2 \right)  \ = \ 2  \left| y_\tau \right|^2 \left( \tilde{m}_{\tilde{L}}^2 + \tilde{m}_{\tilde{e}}^2 + \tilde{m}_{h_d}^2  \right)
	\ + \  2 \left| y_\tau \right|^2 \:  \left( \left| \tilde{\xi}_\tau \right|^2 \ + \   \tilde{\xi}_\tau \mu^* +  \tilde{\xi}_\tau^* \mu \right) \ - \ \frac{3}{5}g_1^2 \mathcal S  \; , \\ &
				\label{eq:b2a_tmL} \\ 
\end{split}\\
\begin{split}
&16\pi^2 \: \beta\left( \tilde{m}_{\tilde{e}_3}^2 \right)  \ = \ 4  \left| y_\tau \right|^2 \left( \tilde{m}_{\tilde{L}}^2 + \tilde{m}_{\tilde{e}}^2 + \tilde{m}_{h_d}^2  \right)
	\ + \  4 \left| y_{\tau} \right|^2 \left( \:  \left| \tilde{\xi}_\tau \right|^2 \ + \    \tilde{\xi}_\tau \mu^* +  \tilde{\xi}_\tau^* \mu \right) \ + \ \frac{6}{5}g_1^2 \mathcal S  \; , \\ &
				\label{eq:b2a_tme} \\ 
\end{split}\\
\begin{split}
&16\pi^2 \: \beta\left( \tilde{m}_{h_u}^2 \right)  \ = \  6 \left| y_t \right|^2 \left( \tilde{m}_{\tilde{q}}^2 + \tilde{m}_{\tilde{u}}^2 + \tilde{m}_{h_u}^2  \right)	
	\ + \  6  \left| y_b \right|^2 \:  \left( \left| \tilde{\xi}_d \right|^2 \ + \   \tilde{\xi}_d \mu^* +  \tilde{\xi}_d^* \mu \right) \\ &
\qquad \qquad 
                      \ + \ 2 \left| y_{\tau} \right|^2 \left(\left|\tilde{\xi}_\tau \right|^2 + \tilde{\xi}_\tau \mu^* +  \tilde{\xi}_\tau^* \mu \right) \ + \ \frac{3}{5}g_1^2 \mathcal S  \; , \\ &
		\label{eq:b2a_tmhu} \\
\end{split}\\
\begin{split}
&16\pi^2 \: \beta\left( \tilde{m}_{h_d}^2 \right) \ = \   6  \left| y_b \right|^2 \left( \tilde{m}_{\tilde{q}}^2 + \tilde{m}_{\tilde{d}}^2 + \tilde{m}_{h_d}^2  \right) 
 \ + \ 2  \left| y_{\tau} \right|^2 \left( \tilde{m}_{\tilde{L}}^2 + \tilde{m}_{\tilde{e}}^2 + \tilde{m}_{h_d}^2  \right)\\ &
                      \qquad \qquad
	\ + \  	6 \left| y_t \right|^2  \: \left( \left| \tilde{\xi}_u \right|^2  \ + \ \tilde{\xi}_u \mu^* +  \tilde{\xi}_u^* \mu \right)  \ - \ \frac{3}{5}g_1^2\mathcal S  
			\label{eq:b2a_tmhd} \, .  \\
\end{split}
\end{align}    

The soft mass RGE must be complemented by the RGE for the Higgs sector parameters, $\tilde{\xi}_u, \tilde{\xi}_d, \tilde{\xi}_{\tau}, \mu$ and $b_\mu$:
\begin{align}
16\pi^2 \: \beta\left( \tilde{\xi}_u \right) \ &=   \left(3 \left| y_t \right|^2 +3 \left| y_b \right|^2 + \left|y_{\tau}\right|^2 \right) \tilde{\xi}_u  \ 
	- \ 2  \left| y_b \right|^2  \left( \tilde{\xi}_u + \tilde{\xi}_d \right) + \tilde\xi_u \left(3 g_2^2+\frac{3}{5}g_1^2\right)   \; ,\label{eq:b2a_txi}     \\
\begin{split}
16\pi^2 \: \beta\left( \tilde{\xi}_d \right) \ &=  \left(3 \left| y_t \right|^2 + 3\left| y_b  \right|^2 +\left|y_{\tau}\right|^2\right) \tilde{\xi}_d  \ 
	- \ 2  \left| y_t \right|^2  \left( \tilde{\xi}_u + \tilde{\xi}_d \right)  \\
& \qquad \qquad  \qquad \qquad 
	+2\left|y_{\tau}\right|^2\left(\tilde\xi_\tau-\tilde\xi_d\right)+\tilde\xi_d \left(3 g_2^2+\frac{3}{5}g_1^2\right)   \; ,
\label{eq:b2a_teta}  
\end{split}	 \\
16\pi^2 \: \beta\left( \tilde{\xi}_\tau \right) \ &=  \left(3 \left| y_t \right|^2 + 3\left| y_b  \right|^2 +\left|y_{\tau}\right|^2\right) \tilde{\xi}_\tau  \ 
	+ \ 6  \left| y_b \right|^2  \left( \tilde{\xi}_d - \tilde{\xi}_\tau \right) +\tilde\xi_\tau \left(3 g_2^2+\frac{3}{5}g_1^2\right)    \; ,\label{eq:b3a_teta}   \\
16\pi^2 \: \beta\left( \mu\right) \ &=   \left( 3 \left| y_t \right|^2 + 3 \left| y_b  \right|^2 + \left|y_{\tau}\right|^2-3 g_2^2-\frac{3}{5}g_1^2 \right) \mu    \; , \\ \label{eq:b2a_mu}
16\pi^2 \: \beta\left( b_\mu\right) \ &=   \left( 3 \left| y_t \right|^2 + 3 \left| y_b  \right|^2+\left| y_{\tau}\right|^2-3g_2^2-\frac{3}{5}g_1^2 \right) b_\mu  .
\end{align}
The RGEs of Dirac gauginos and Majorana adjoint fermions can be found in~\cite{Goodsell:2012fm}. Examining these RGE, there are several features worth mentioning. First, the first and second generation squarks and sleptons can be found by setting the Yukawa couplings in Eq.~\eqref{eq:b2a_tmq}-\eqref{eq:b2a_tme} to zero. As the traditional gaugino mediated contribution to the soft mass RGEs is absent because of the Dirac nature of our gauginos, the first and second generation squarks/sleptons only renormalize due to hypercharge $D$-term. Written compactly,
\begin{align}
16\pi^2 \: \beta\left( \tilde{m}_{\phi}^2 \right) \ = \frac{6}{5}Y_{\phi}\, g^2_1\, \mathcal S,
\label{slight}
\end{align}
where $\phi$ is a first or second generations sfermion with hypercharge $Y_{\phi}$. A second feature of these RGE is the unusual piece proportional to $\xi_u \mu^*$ seen in e.g. Eq.~\eqref{eq:b2a_tmq}. As shown in Fig.~\ref{fig:first}, this piece can be traced to insertions of trilinear scalar term from $\mu$-operator and the $\xi$-operator. 

The final ingredient needed to complete the RGE for this theory is the running of $\mathcal S$:
\begin{equation}
\begin{split}
16\pi^2 \: \beta \left(\mathcal{S} \right) \ & = \ \frac{66}{5} g_1^2 \mathcal{S} 
	\ - \ 12 \left|y_t\right|^2\left(\left|\tilde\xi_u\right|^2 + \tilde\xi_u\mu^*+\tilde\xi_u^*\mu\right)   \\
& \qquad \qquad 	+ \  12 \left|y_b\right|^2 \left( \left|\tilde\xi_d\right|^2+\tilde\xi_d\mu^*+\tilde\xi_d^*\mu \right) \ + \ 
	4 y_{\tau}^2\left(\left|\tilde\xi_d\right|^2+\tilde\xi_d\mu^*+\tilde\xi_d^*\mu\right)    \;.
\end{split}
\label{eq:b2a_sr}
\end{equation}
In the limit the  $\tilde{\xi}$ parameters are taken to zero, Eq.~\eqref{eq:b2a_sr} reduces to its conventional MSSM form.

\section{Solutions}
\label{sec:4}
\begin{figure}[h]
\centering
\includegraphics[width=0.80\textwidth]{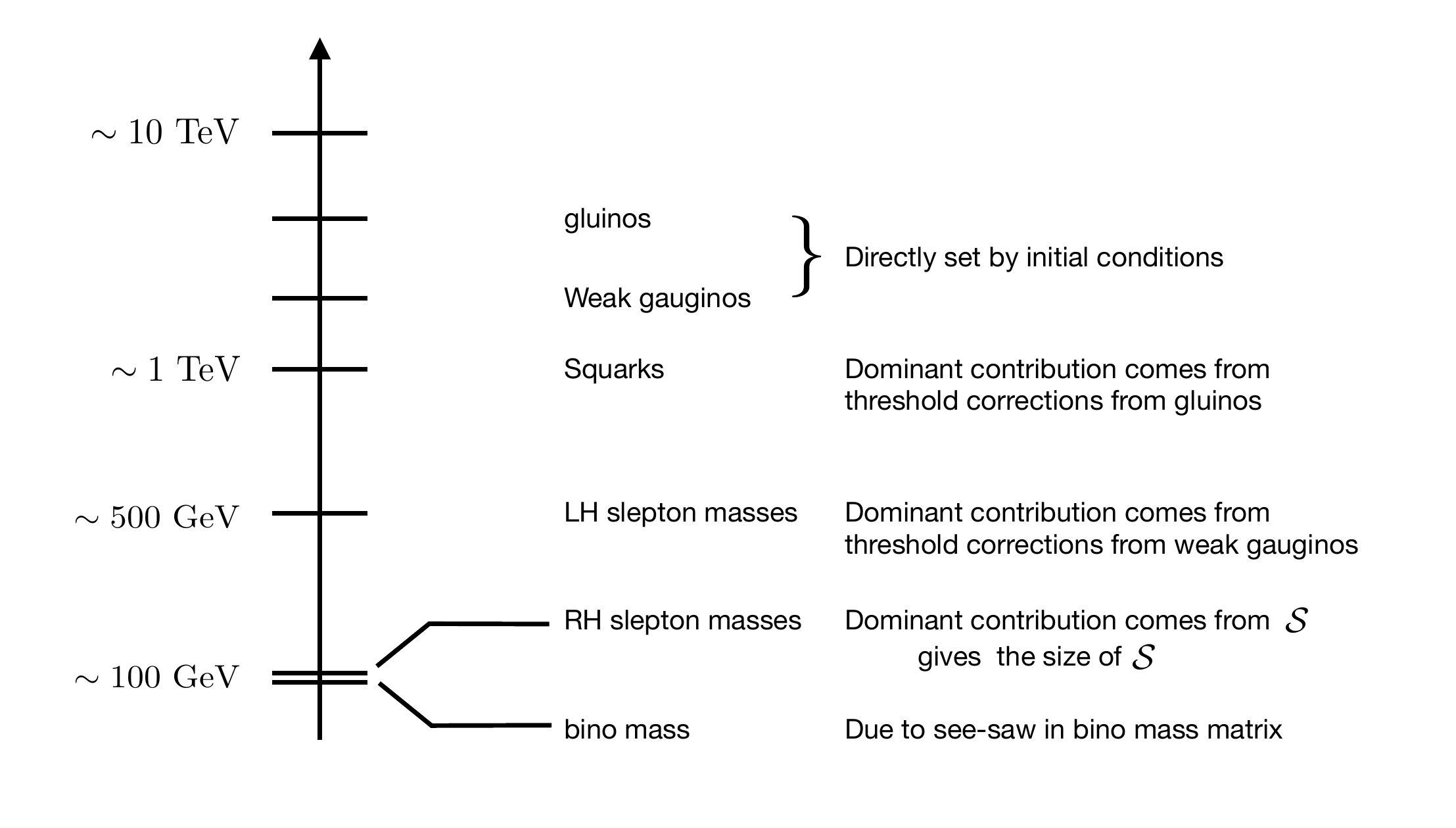}
\caption{A schematic diagram of the spectrum to show various scales.}
\label{fig:schematic}
\end{figure}

Our objective in this section is to determine the spectra at $\MuIR$ and other parameters after solving equations listed in Eqs.~(\ref{eq:b2a_txi}-\ref{eq:b2a_tmhd}) using all the initial conditions specified at the boundary $\MuInt$ (Eq.~\eqref{eq:input-toy}). Before we proceed any further, however, we give the schematics of the scales we target. We think that having a prior understanding of the scales (i.e., sizes of various terms) allows one to visualize and consequently appreciate the solutions, especially the analytical part, better.  We have plotted a schematic representation of superpartner mass spectrum in Fig.~\ref{fig:schematic}. Here is a simplified summary of the mass scales. 

The key spectral features are:
\begin{itemize}

\item As expected, the lowest lying scale represents the mass of the LSP. For the relic abundance to work out we rely on co-annihilation of the LSP (mostly bino) with RH sleptons. This scale, therefore, also represents masses of the RH sleptons. We take this opportunity to reiterate that gaugino mediated contribution (loop suppressed and finite) from bino can not generate this mass-scale.  In this work, we generate this scale by the hypercharge $D$-term $\mathcal{S}$. The mass of dark matter therefore also directly gives the size of the $S$-term.  

\item Decays of left handed sleptons to LSP give rise to hard leptons and consequently these need to be significantly heavier than the LSP mass scale (or the scale of RH sleptons). Setting the LH sleptons above the LHC bound, therefore, gives rise to a second scale in the spectrum. Masses for the LH sleptons are generated by finite corrections from Wino masses. This allows us, in-turn, to set the scale for wino mass.  

\item Finally, LHC bounds on colored particles imply that squarks need to be significantly heavier.  The primary source for squark masses are finite corrections from gluinos. Setting the squark masses to be around the TeV scale, one then finds masses for gluinos. 

\item The final piece is the mass scale of the higgsinos. This can be determined either by Dark matter direct detection constraints which limits the higgsino fraction in the lightest bino-like neutralino or by direct LHC searches. At LHC, heavy higgsinos can decay to the LSP associated with Higgs or Z-boson. The non-observation of such events puts an upper bound on the higgsino masses.

\end{itemize}

\subsection{Analytical solutions:}
It is clear that, even at one loop order, we need to solve the RGE Eq.~\eqref{eq:b2a_tmq}-\eqref{eq:b2a_sr}  numerically. However, in order to develop some intuition for the gross features of the spectrum, we start by making some simplifying assumptions which allow us to solve the RGE analytically. As we show later in this subsection, most of the phenomenological aspects of this work, such as finding a viable candidate for Dark Matter or the spectrum of colored particles, can be understood within this simplified picture. Calculating the Higgs mass, however, requires more careful considerations and will only be discussed in the context of full numerical solutions. 

To simplify the RGE, in the following subsections we will ignore all Yukawa couplings except for the top Yukawa $y_t$. Even though we use non-zero $\mathcal{S}_0$, we will use $\tilde{\xi}_0 = 0$, which ensures that none of the $\xi$ operators will play a role. While this approximation may seem unjustified given our initial conditions, we find that the full, numerical solution derived later is well approximated by the results we derive with $\tilde{\xi}_0 = 0$.

\subsubsection{First and second generation sfermions:}
\label{sec:first_sec_sfermions}

As mentioned earlier, the first and second generation sfermions only run because of the hypercharge $D$-term $\mathcal S$. With the Yukawa couplings zeroed, the running of $\mathcal S$ is easy to work out:
\begin{align}
\mathcal{S} (\MuIR) \ &= \ \mathcal{S}_0 \ \left[  \frac{g_1(\MuIR)}{g_1(\MuInt)} \right]^{66/5b_1}, 
\label{eq:srun} \; 
\end{align}
where $b_1$ is the beta function for hypercharge and we have imposed the boundary conditions from Eq. \eqref{eq:scalar_bound}. Plugging Eq.~\eqref{eq:srun} into the sfermion RGE (Eq.~\eqref{slight}), we find
\begin{align}
\tilde m^2_{\phi} (\MuIR)=  -\frac{6}{5}\, q_Y\, \mathcal{S}_0\, \frac{\alpha_1(\MuIR)}{4\pi}  \log \left( \frac{\MuInt}{\MuIR} \right),
\label{eq:srun2}
\end{align} 
 where $q_Y$ is the hypercharge of the sfermion. No matter what sign we choose for $\mathcal{S}_0$, this (radiative) mass squared will be negative for some matter fields ($ \tilde{q}_i, \tilde{u}_i, \tilde{d}_i, \tilde{l}_i, \tilde{e}_i $), simply because they don't all have the same sign hypercharge. If Eq.~(\ref{eq:srun2}) were the only contribution to the sfermion masses, this result would be fatal as charge/color breaking minima would occur. Fortunately, as illustrated in Fig.~\ref{fig:schematic}, the LH sfermion masses receive positive definite and finite contributions from loops of gauginos~\cite{Fox:2002bu}. As Eq.~(\ref{eq:srun2}) is proportional to the hypercharge coupling and only logarithmically sensitive to $\MuInt$, it is entirely possible for the finite contribution proportional to $\alpha_2(\MuIR),\, \alpha_3(\MuIR)$ to dominate over Eq.~\eqref{eq:srun2}. This logic suggests that we should choose $\mathcal S_0 < 0$, so that Eq.~(\ref{eq:srun}) is positive for the right handed sleptons. For example, if $\MuInt=10^{11}$ GeV we need $\left|\mathcal S_0\right|\sim \left(700~\text{GeV}\right)^2$ in order to achieve 100 GeV right handed slepton masses. Such a value of $\mathcal S$ can be obtained by properly choosing UV inputs $\mu_u^0$ and $\mu_d^0$. This choice for $\mathcal S_0$ renders the contribution of Eq.~(\ref{eq:srun2}) to $\tilde m^2_{\tilde q}$ and $\tilde m^2_{\tilde d}$ also positive, while $\tilde m^2_{\tilde{\ell}}$ and $\tilde m^2_{\tilde u}$ receive a negative contribution.  As we will show, negative $\tilde m^2_{\tilde{\ell}}, \tilde m^2_{\tilde u}$ must be offset by the finite wino and gluino loops, and the requirement that $\tilde m^2_{\tilde{\ell}}(\MuIR) > 0, \tilde m^2_{\tilde u}(\MuIR) > 0$ can be used to restrict the input wino/gluino mass parameters. In order to impose this restriction, we first need to know how the $\MuIR$ gaugino masses depend on the $\MuInt$ inputs. 

\subsubsection{Colored Sector:}
\label{sec:colored}
The gluino sector (gluino + adjoint partner) masses at $\MuIR$ are straightforward to calculate. Using the boundary conditions set in Sec.~\ref{sec:4}, the renormalized masses are. 

\begin{align}
\setlength{\jot}{10pt}	
&	\alpha_s \left( \MuInt \right)   \ =  \ \alpha_s \left(\MuIR \right) \; , \label{eq:4.1.1}\\
&	Z_{\Sigma_3} \left( \MuIR \right)  \ = \ 
			Z_{\Sigma_3} \left(  \MuInt \right) \left( \frac{\mu}{\Lambda}\right)^{\frac{3 \alpha_s}{\pi}} \; , \label{eq:4.1.2} \\
&	M_{D_3} \left( \MuIR \right)   \ =  \ \sqrt{\frac{\alpha_s \left( \MuIR \right)}{\alpha_s \left(\MuInt \right)} }  
			\sqrt{\frac{Z_{\Sigma_3} \left( \MuInt \right)}{Z_{\Sigma_3} \left(\MuIR \right)}}  \ M_{D_3} \left( \MuInt \right) = M_{D_3} \left( \MuInt \right)\,\left( \frac{\mu}{\Lambda}\right)^{-\frac{3 \alpha_s}{2\,\pi}}\;,
		  \label{eq:4.1.3}\\[3ex]
&	M_{\Sigma_3} \left( \MuIR \right)  = \frac{Z_{\Sigma_3} \left(\MuInt \right)}{Z_{\Sigma_3} \left( \MuIR \right)}  \ M_{\Sigma_3} \left( \MuInt \right) =   M_{\Sigma_3} \left( \MuInt \right)\,\left( \frac{\mu}{\Lambda}\right)^{-\frac{3 \alpha_s}{\pi}}, \label{eq:4.1.4}	
\end{align}
where $Z_{\Sigma_3}$ is the field strength renormalization of the color adjoint and Eq. \eqref{eq:4.1.1} contains the well-known result that the QCD gauge coupling does not run with the supersoft field content. Using $\psi_3$ to denote the fermion within $\Sigma_3$, the color adjoint fermion masses can be written in matrix form as
\begin{equation}
	 	\begin{pmatrix} \tilde{g} & \psi_3 \end{pmatrix}  
		\begin{pmatrix} 0 & M_{D_3} \\ M_{D_3} & M_{\Sigma_3}  \end{pmatrix} 
		\begin{pmatrix}  \tilde{g} \\ \psi_3 \end{pmatrix}.    \label{eq.4.1.5}
\end{equation}
Depending on the relative strength of the Dirac mass of gluino ($M_{D_3}$) and the Majorana mass of $\psi_3$ ($M_{\Sigma_3}$), three qualitatively distinct spectrum at the scale $\MuIR$ emerge. i.) primarily Majorana gluinos, ii.) primarily Dirac gluinos, and iii.) mixed Majorana-Dirac gluinos. For the coupling structure and some collider implications of the different possibilities, see Ref.~\cite{Kribs:2013eua}.

The gluino mass matrix can acquire all the three types of textures even if  $M_{\Sigma_3}$ and $M_{D_3}$ start out being equal at the messenger scale (depending on whether $\Sigma_3$ self interactions are present or not). It is instructive to properly diagonalize the mass matrix and write the effective theory in terms of mass eigenstates 
\begin{align}
	\begin{pmatrix} \tilde{g}_l \\ \tilde{g}_h \end{pmatrix}  \ = \ 
	\begin{pmatrix} \cos \theta_g & \sin \theta_g \\ - \sin \theta_g & \cos \theta_g  \end{pmatrix} 	
	\begin{pmatrix} \tilde{g} \\ \psi_3 \end{pmatrix}  \\ 
	\cos^2 \theta_g \ = \ \frac{1}{2} \left(  1 - \frac{M_{\Sigma_3}}{\sqrt{M_{\Sigma_3}^2 + 4 M_{D_3}^2} } \right) \;,	
\end{align}
In the above,  $\tilde{g}_l \text{ and } \tilde{g}_h$ represent the light and heavy gluino spinors with masses $M_{g_l}$ and $M_{g_h}$ respectively. 
\begin{equation}
\begin{split}	
	\mathcal{L} \  \supset  \ & \frac{1}{2} M_{g_l} \tilde{g}_l \tilde{g}_l  + 
			\frac{1}{2}  M_{g_h} \tilde{g}_h \tilde{g}_h \;, \\
       - M_{g_l}   \  = \ \frac{1}{2} \left(  M_{\Sigma_3}   - \sqrt{M_{\Sigma_3}^2 + 4 M_{D_3}^2} \right) & \quad \text{and} \quad
	 M_{g_h} \  = \  \ \frac{1}{2} \left(  M_{\Sigma_3}   + \sqrt{M_{\Sigma_3}^2 + 4 M_{D_3}^2} \right) \;.
\end{split}
\label{eq:gluino_mass}
\end{equation}
This decomposition is valid even if gluinos are purely Dirac; there, $M_{\Sigma_3} = 0$, $M_{g_l} = - M_{g_h}$, and  $\sin \theta_g  = \cos \theta_g  = 1/\sqrt{2}$. 

One of the important features of the generalized supersoft spectrum is that -- regardless of the Majorana/Dirac composition of the gluino in all these three cases -- squark masses remain supersoft, i.e. the gluino mediated squarks masses do not pick up any $\log \MuInt$ sensitivity.  Below, we verify this statement  using the diagrams in Fig.~\ref{fig:third}: 

\begin{figure}[h]
\centering
\includegraphics[width=0.7\textwidth]{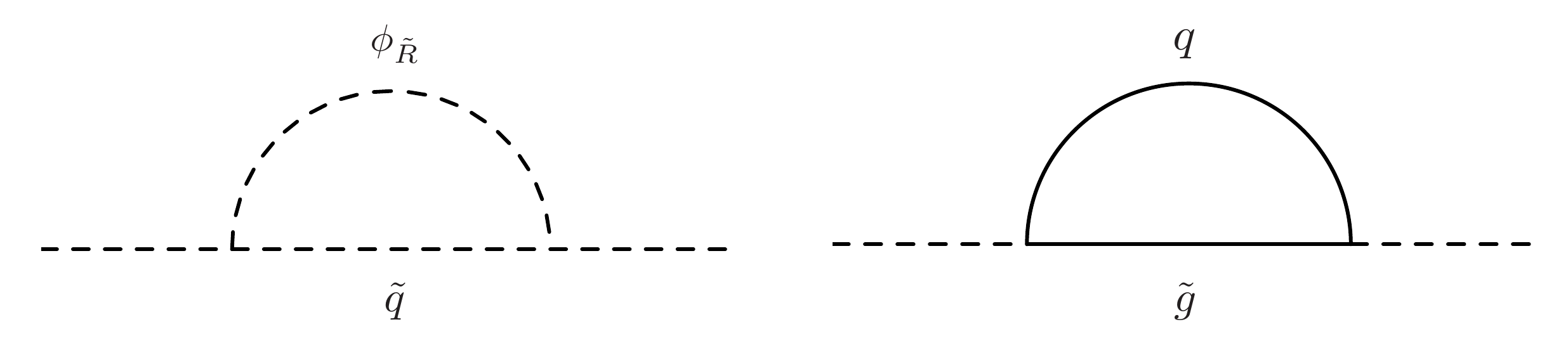}
\caption{Gaugino mediated masses of the squark fields. The purely scalar loop cancels the logarithmic divergence which appears in the prototypical gaugino mediated correction to squark masses.}
\label{fig:third}
\end{figure}

 \begin{equation}
\begin{split}	
	 \tilde m_{q}^2  \ & = \  - \frac{\alpha_s}{\pi} \: \text{C}_2 \left( r \right) \: 
		\int \! \! d^4 k  \: \left( \frac{\cos^2 \theta_g}{k^2 - M_{g_l}^2} +
			\frac{\sin^2 \theta_g}{k^2 - M_{g_h}^2} \right)  + 
				\frac{\alpha_s}{\pi} \: \text{C}_2 \left( r \right) \: 
		\int \! \! d^4 k  \: \frac{M_{D_3}^2}{k^2 (k^2 - m_{\phi_R}^2)}  \\
		& \supset \  - \frac{\alpha_s}{\pi}  \: \text{C}_2 \left( r \right) \: 
			\left(   \cos^2 \theta_g  M_{g_l}^2 + \sin^2 \theta_g M_{g_h}^2 \right)  \log \MuInt^2 
			+ \frac{\alpha_s}{\pi}  \: \text{C}_2 \left( r \right) \:  M_{D_3}^2 \log \MuInt^2 \;.
\end{split}		
\label{eq:deltam}
\end{equation}
Here, $\tilde m^2_q$ is the finite correction squark mass generated generated at $\MuIR$ and $C_2(r)$ is the quadratic casimir. Examining Eq.~\eqref{eq:deltam}, the first integral is due to $g_l$ and $g_h$ running in the loop, and the second integral is due to the real part of the scalar octet of mass $m_{\phi_R}$. The $\log \MuInt^2$ term is cancelled between the two terms, as
\begin{equation}
	 \cos^2 \theta_g  M_{g_l}^2 + \sin^2 \theta_g M_{g_h}^2  \ = \ - M_{g_l} M_{g_h} \ = \ M_{D_3}^2 \;.
\end{equation}
Cleaning up Eq.~\eqref{eq:deltam}, the gluino-induced contributions to the soft masses (squared) of the squarks  ($\tilde{m}_{\tilde{q}}^2$,\, $\tilde{m}_{\tilde{u}}^2$,\, $\tilde{m}_{\tilde{d}}^2$) can be written in terms of the mass eigenvalues and the gluino mixing angle, 
\begin{equation}
	 \tilde{m}_{\tilde{q}}^2 \ = \ \frac{\alpha_s}{\pi}  \: \text{C}_2 \left( r \right) \: \bigg \{
		M_{D_3}^2 \log \frac{m_{\phi_3}^2}{M_{D_3}^2} + M_{D_3}^2 \cos 2 \theta_g  \log \left( \tan^2 \theta_g \right) \bigg \}  
		\;. \label{eq:4.1.11}
\end{equation}
where $m_{\phi_3}$ is the mass of the scalar adjoint component in $\Sigma_3$. At tree level, $m_{\phi_3} = 2 M_{D_3}$, though running and the existence of Majorana masses will change the relationship somewhat. 

If we assume that finite gluino contribution represents the squark masses, we can compare Eq.~\eqref{eq:4.1.11} to the LHC bounds on colored sparticles to get an idea for the allowed ranges of UV inputs $M^0_{D_3}, M^0_{\Sigma_3}$. The IR masses for the squarks and lightest gluino sector are shown below in Fig.~\ref{fig:coloredstuff} as a function of the input masses. 
\begin{figure}[h!]
\centering
\includegraphics[width=\textwidth]{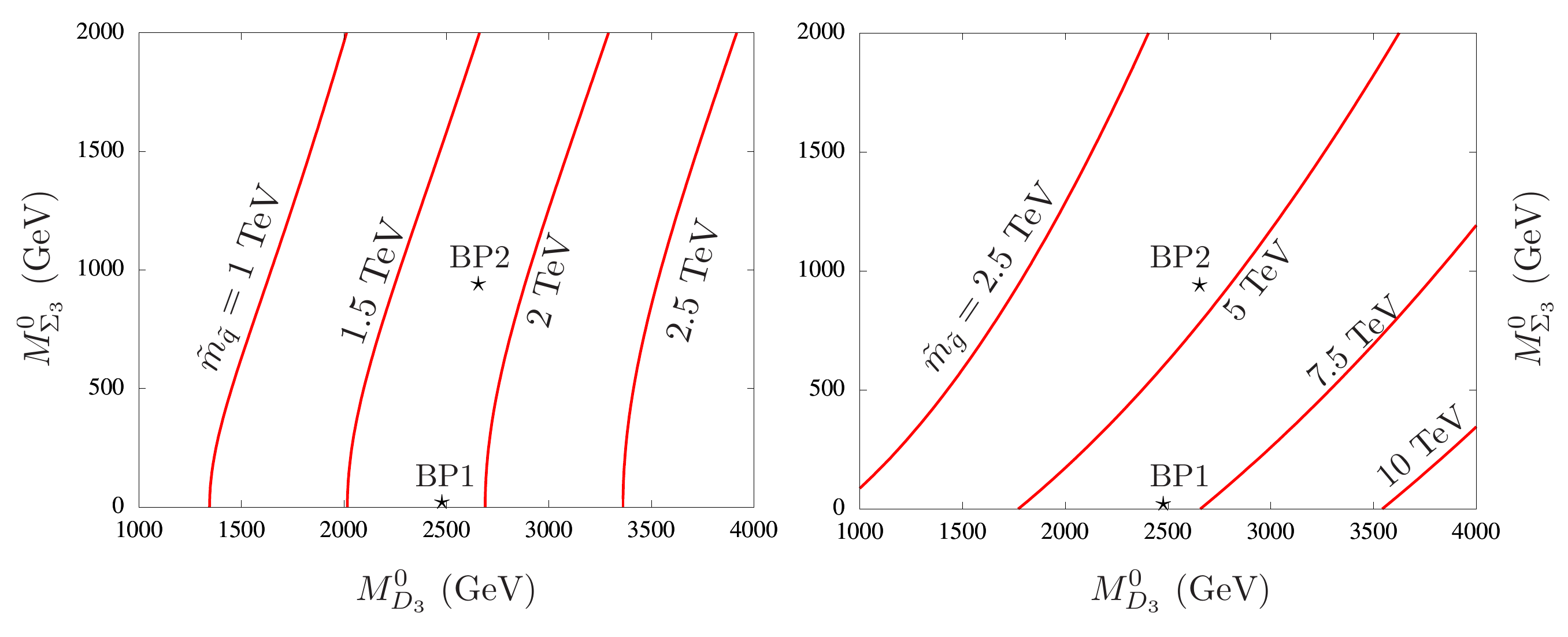}
\caption{Left: finite correction to the squark mass $\tilde m_{\tilde q}(\MuIR)$ as a function of the UV inputs: the Dirac mass $M^0_{D_3}$ and adjoint Majorana mass $M^0_{\Sigma_3}$. Right: mass of the lightest eigenvalue of the gluino sector as a function of the same UV inputs. For both plots, we used the tree level relation $m_{\phi_3} = 2 M_{D_3}$ for simplicity and took $\MuIR = 10^3\, \text{GeV}$, $\MuInt = 10^{11}\, \text{GeV}$ and  $\alpha_3(\MuIR) = \alpha_3(\MuInt) = 0.118$ as inputs. The starred point $(M^0_{D_3}, M^0_{\Sigma_3} ) = (2.48\, \text{TeV}, 0\, \text{TeV})$ for BP1 and $(M^0_{D_3}, M^0_{\Sigma_3} ) = (2.65\, \text{TeV}, 940\, \text{GeV})$ for BP2, yields $\tilde m_{\tilde q} \simeq 1.8\, \text{TeV}$ which satisfy the present LHC bounds~\cite{Aaboud:2017iio,Sirunyan:2017cwe}, and a gluino mass of $7\, \text{TeV}$ and $4.6~\text{TeV}$ respectively.}
\label{fig:coloredstuff}
\end{figure}
As the squark masses are radiatively generated and finite, the squarks are significantly lighter than the gluinos for a given set of inputs. The shape of the squark mass curves can be understood from the fact that the Majorana gluino sector mass runs significantly faster than the Dirac mass. For $M^0_{\Sigma_3} \sim M^0_{D_3}$, $M_{\Sigma_3}(\MuIR) \gg M_{D_3}(\MuIR)$, effectively, the gluino is Majorana like with eigenvalue $M_{D_3}^2/M_{\Sigma_3}$. In such a scenario, the log emerges in the gluino mediated correction to the scalar masses but with a cut off $M_{\Sigma_3}$, i.e., $\text{Log}\left(M_{\Sigma_3}/M_{D_3}\right)$. In assuming the squark masses are set by Eq.~\eqref{eq:4.1.11}, we have ignored i.) finite (positive) contributions from loops of $SU(2)$ or $U(1)$ gauginos, ii.) the log-enhanced hypercharge $D$-term contribution proportional to $\mathcal S_0$. While necessary for getting the exact spectrum of the theory, these contributions are both subdominant to Eq.~\eqref{eq:4.1.11}; the $SU(2)$ and $U(1)$ gaugino loops are suppressed by the smaller EW couplings, and the $\mathcal S_0$ contribution is small because, as explained in Sec.~\ref{sec:first_sec_sfermions}, it sets the mass of the lightest sfermions. Therefore, our assumption that Eq.~\eqref{eq:4.1.11} sets the squark mass is justified, and we can use Fig.~\ref{fig:coloredstuff} to rule out $M^0_{D_3} \lesssim 2\, \text{TeV}$. These bounds are rough, as the detailed phenomenology will depend on how `Dirac-like' vs. `Majorana-like' the lightest gluino eigenstate is; see Ref.~\cite{Kribs:2012gx,Kribs:2013eua}.

\subsubsection{Electroweak Sector:}
\label{sec:ew}
Following the same procedure as above, we can calculate the electroweak gaugino masses (both Dirac and Majorana pieces). One difference in the electroweak case is that the gauge couplings do run in a supersoft theory, with beta function coefficients $b_1=33/5$ and $b_2=3$. Working in the $y_t \to 0$ limit, the masses at the $\MuIR$ scale are:
\begin{align}
\setlength{\jot}{10pt}	
	\frac{1}{\alpha_i \left( \MuInt \right) }  \ & =  \frac{1}{\alpha_i \left( \MuIR \right) } - \frac{b_a}{2 \pi} \log \left( \frac{\MuInt}{\MuIR}\right) \;,
			\label{eq:4.2.1}  \\
	Z_{\Sigma_a}  \left( \MuIR \right)  \ & = \ Z_{\Sigma_a} \left( \MuInt \right)  
			\left( \frac{\alpha_i \left( \mu_R \right) }{\alpha_i \left( \MuInt \right)}
				\right)^{\frac{2\text{C}_2 (\text{Adj})}{b_a}}  \;, \label{eq:4.2.2} \\ 
	M_{D_a} \left( \MuIR \right)   \ & =  \ M_{D_a}^0   \: 
			\left( \frac{\alpha_a \left( \MuIR \right) }{\alpha_i \left( \MuInt \right)}
				\right)^{\frac{1}{2} - \frac{\text{C}_2 (\text{Adj})}{b_a}}  \;, \label{eq:4.2.3} \\
	M_{\Sigma_a} \left( \MuIR \right)   \ & = \ M_{\Sigma_i}^0   \:
			\left( \frac{\alpha_i \left( \MuIR \right) }{\alpha_i \left( \MuInt \right)}
				\right)^{- \frac{2\text{C}_2 (\text{Adj})}{b_a}}    \;,	\label{eq:4.2.4}
\end{align}
where $a = 1,2$ and $C_2(\text{Adj})$ is the quadratic Casimir of the adjoint representation. 

Focusing first on the $SU(2)$ sector, loops of winos and $SU(2)$ adjoints generate a finite correction to the mass of all $SU(2)$ charged sfermions. The contribution has the same form as Eqs.~\eqref{eq.4.1.5}-\eqref{eq:4.1.11}, but with $M_{D_2}$ replacing $M_{D_3}$, $\alpha_2$ replacing $\alpha_3$, and $\theta_2$ replacing $\theta_g$:
\begin{equation}
	 \tilde{m}_{\tilde\ell}^2 \ = \frac{\alpha_2}{\pi}  \: \text{C}_2 \left( r \right) \: \bigg \{
		M_{D_2}^2 \log \frac{m_{\phi_2}^2}{M_{D_2}^2} + M_{D_2}^2 \cos 2 \theta_2  \log \left( \tan^2 \theta_2 \right) \bigg \}  
		\;. \label{eq:4.1.18}
\end{equation}
This effect is particularly important for the left-handed sleptons, as the contribution to their masses from the $\mathcal S_0$ piece is negative (see discussion following Eq.~\eqref{eq:srun2}) and they receive no contribution from gluino loops. In fact, if we follow the same logic as in the previous section and neglect the $\mathcal S_0$ and bino loop terms, the left handed slepton mass is completely set by Eq.~\eqref{eq:4.1.18}. The slepton mass and the lightest wino sector eigenvalue are shown below in Fig.~\ref{fig:su2stuff} as a function of the UV inputs $M^0_{D_2}, M^0_{\Sigma_2}$.
\begin{figure}[h!]
\centering
\includegraphics[width=\textwidth]{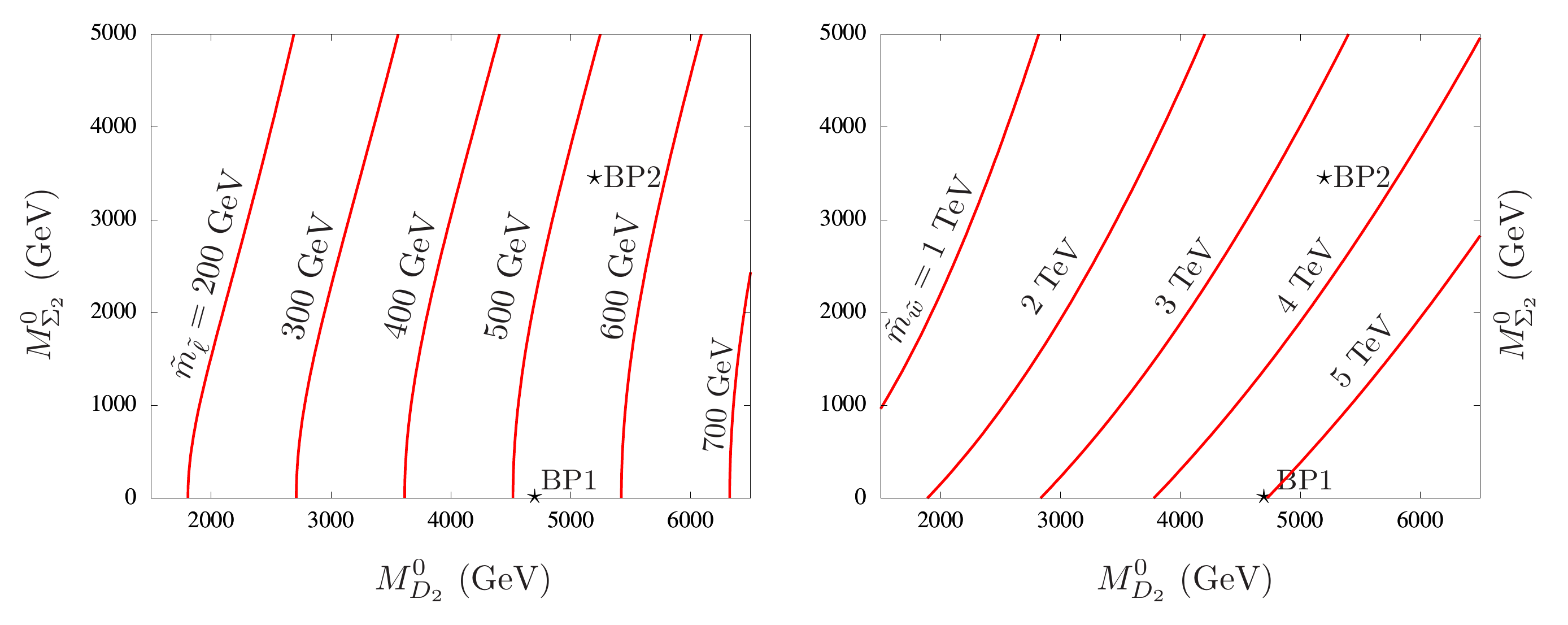}
\caption{Left: finite contribution to all $SU(2)$ doublet sfermion masses, $\tilde m_{\tilde{\ell}}(\MuIR)$ as a function of the UV wino-sector masses. Right: lightest wino-sector ($2\times 2$) mass eignenvalue after running. For both plots, $\MuIR = 10^3\, \text{GeV}$, $\MuInt = 10^{11}\, \text{GeV}$ and  $\alpha_2(\MuIR) = 0.033$. The starred point $(M^0_{D_2} , M^0_{\Sigma_2})= (4.7\, \text{TeV}, 0\, \text{TeV}$) for BP1 and  $(M^0_{D_2} , M^0_{\Sigma_2})= (5.2\, \text{TeV}, 3.45\, \text{TeV}$) for BP2 yields a finite mass contribution of $565\, \text{and}\, 515\, \text{GeV}$ respectively for all $SU(2)$ doublets, and wino eigenvalue of $5\, \text{TeV}$ and $3.4\,\text{TeV}$ respectively.}
\label{fig:su2stuff}
\end{figure}

We can use Fig.~\ref{fig:su2stuff} to get a rough idea of what range of UV inputs are allowed. For sleptons significantly heavier than the LSP, the current limits are $\sim 500\, \text{GeV}$~\cite{Aad:2014vma,Aad:2015baa} from the process $pp\to \tilde{\ell}^+\tilde{\ell}^-, \tilde{\ell}^{\pm} \to \ell^{\pm}+\text{LSP}$. Imposing this constraint selects $M^0_{D_2} > 5.2\,\text{TeV}$, however the situation is a bit more subtle. One complication is that the mass eigenstate electroweakinos are actually a combination of wino sector fields, bino sector fields, and Higgsinos, and we have so far neglected mixing among these different multiplets. The full mass matrix for the electroweak gauginos is given in Appendix~\ref{appendix:A} and will be used in Sec.~\ref{sec:bench} when we present benchmark points. The second issue is that the Higgs multiplets are also charged under $SU(2)$ so their soft mass receives a boost from Eq.~\eqref{eq:4.1.18}. However, to achieve EWSB, we need one Higgs mass (squared) eigenvalue to become negative.  Large, positive contributions to $m^2_{H_u}, m^2_{H_d}$ introduce some tension into the Higgs mass system, since they must be countered by other contributions to the soft masses or by a large $b_{\mu}$ term. Both options introduce some degree of tuning. Because of these complications, an accurate description of the allowed region in $M^0_{D_2}, M^0_{\Sigma_2}$ space will need to wait until we study EWSB and the Higgs mass, Sec.~\ref{sec:mH}.

Turning to the $U(1)_Y$ (bino) sector, in the limit $\langle H_u \rangle = \langle H_d \rangle = 0$ the bino sector mass matrix has the same form as the gluino sector in Eq.~\eqref{eq.4.1.5} with the substitutions $M_{D_3} \rightarrow M_{D_1}$ and $M_{\Sigma_3} \rightarrow M_{\Sigma_1}$. While any of the Dirac/Majorana mass hierarchies mentioned in the gluino section are theoretically possible for the bino, we would like the lightest bino sector eigenstate to play the role of Dark Matter. As such, the lightest bino eigenvalue should be predominantly Majorana in order avoid the stringent direct detection constraints on electroweakly charged Dirac Dark Matter~\cite{Hsieh:2007wq}. A predominantly Majorana bino is achieved by taking $M_{\Sigma_1} \gg M_{D_1}$ (the `see-saw' limit), in which case
\begin{equation}
 M_\text{DM} \ = \ \text{smallest eigenvalue of bino sector mass matrix}   \ \approx \ \frac{M_{D_1}^2} {M_{\Sigma_1}}  
 	+ \mathcal{O} \left( \frac{v^2}{\mu} \right).
	\label{eq:MDM}
\end{equation}
For any desired bino Dark Matter mass, Eq.~\eqref{eq:MDM} fixes this combination of the $\MuIR$ masses $M_{\Sigma_1}, M_{D_1}$, which can be translated into a constrain on the UV parameters $M^0_{\Sigma_1}, M^0_{D_1}$. Unfortunately, the range of masses where an isolated, predominantly Majorana bino Dark Matter particle can satisfy the observed thermal relic abundance is relatively small \cite{ArkaniHamed:2006mb}. However, as discussed around Fig.~\ref{fig:schematic}, the range of viable masses becomes much larger if the lightest right handed slepton has nearly the same mass as the Dark Matter, as coannihilation enhances the annihilation cross section. Within our setup, the condition $\tilde m^2_{\tilde e} \approx M^2_{DM}$ boils down to a relationship among $\MuIR$ masses, which can be transmuted into constraints on the UV parameters. Explicitly, let us approximate the right handed slepton masses by their $\mathcal S_0$ dependent piece, Eq~\eqref{eq:srun}.\footnote{Comparing Eq~\eqref{eq:srun} to the finite contribution from bino-loops (Eq.~\eqref{eq:4.1.11}, with $M_{D_a} \to M_{D_1}$ and $m_{\phi_R} \to m_S$ (the mass of the bino-partner's scalar component), the latter contribution to $\tilde m_{\tilde e}$ is suppressed in the limit $M_{\Sigma_1} \approx m_S \gg M_{D_1}$.} Then, the coannihilation requirement $\tilde m_{\tilde e} \sim M_{DM}$ can be turned into a condition on $\mathcal S_0 \log(\MuInt/\MuIR)$
\begin{align}
\tilde m^2_{\tilde e} = M^2_{DM} &\longrightarrow\,  \mathcal S_0  \log \left( \frac{\MuInt}{\MuIR} \right) = -\frac{6}{5}\frac{M^4_{D_1}}{M^2_{\Sigma_1}} \frac{4\pi}{\alpha_1(\MuIR)} \;.
\label{eq:colim}
\end{align}

This line of thinking can actually be extended further.  Once we turn on $v_u, v_d \neq 0$, the lightest neutralino will contain traces of Higgsino, and a neutralino-neutralino-Higgs interaction. This interaction will mediate spin-independent neutralino-nucleus scattering and can come into conflict with Dark Matter direct detection bounds if the interaction strength is too large~\cite{Akerib:2016vxi}. These problematic interaction vanish in the limit $\mu\gg v_u, v_d$ when we decouple the Higgsinos. Therefore we can turn the direct detection constraint into lower bound on $\mu$. In order to derive the lower bounds on $\mu$, we impose the direct detection constraint by requiring $\mu$ to be sufficiently large compared to the Dark Matter mass:
\begin{equation}
\left| \mu \right|^2  \geq k_{dd} \times M_\text{DM}^2 \ \approx \  - k_{dd} \times  \frac{6}{5} \: \frac{\alpha_1}{4\pi}  \: \mathcal{S}_0 \ \log \left( \frac{\MuInt}{\MuIR} \right)\;.
\end{equation}
where $k_{dd}$ is a constant that contains how $\mu(\MuIR)$ translates to a neutralino-neutralino-Higgs interaction strength and what interaction strength is permitted by current direct detection experiments. Notice that the right hand side of the above equation has the $\log$ in the leading order, whereas $\log$ appears in the left hand side only at the subleading order.  Keeping the dominant terms on both sides of the equation and swapping $\mathcal S_0, \mu$ for $\mu^0_u, \mu^0_d$, we can get an approximate constraint equation:
\begin{equation}
\frac{\left( \mu_u^0 + \mu_d^0\right)^2}{\left|\mu_d^0 \right|^2 - \left|\mu_u^0 \right|^2} \ \gtrsim \ k_{dd}\frac{6}{5} \times \frac{\alpha_1}{\pi}    \log \left(\frac{\MuInt}{\MuIR}\right).
\end{equation} 
This forces us to take $\left|\mu_d^0 \right|^2 \gg \left|\mu_u^0 \right|^2$, an appropriate limit given that we know from Sec.~\ref{sec:first_sec_sfermions} that $\mathcal S_0$ must be negative, further simplification happens,
\begin{equation}
  \log\left(\frac{\MuInt}{\MuIR}\right)  \ \lesssim \  \frac{1}{k_{dd}} \: \frac{5\pi  }{6  \alpha_1 } 
\quad  \Rightarrow  \quad 
 \log_{10}\left(\frac{\MuInt}{\MuIR}\right)  \ \lesssim \  \frac{70}{k_{dd}} \;.
\label{eq:logsize}
\end{equation} 
For example, if $k_{dd} = 10$ and $\MuIR = 1\tev$, we find $\MuInt \lesssim  10^{10}\gev$. A precise limit on $k_{dd}$ could either come from direct detection DM results or from LHC. For example, the present bound on Higgsino mass is roughly 600 GeV for a 100 GeV LSP neutralino. Moreover, DM direct detection experiments such as LUX, puts a constraint on the Higgsino fraction of the lightest DM candidate. For a 100 GeV bino-like DM, Higgsino mass need to be around 500 GeV to evade the direct detection bounds in the paradigm of MSSM. Therefore, the sectors roughly signals to $k_{dd}>5$.  However, In all the above derivations, we have neglected the role of Yukawa couplings. A closed solution of $\mathcal S$ in the presence of the Yukawa couplings is given in Appendix~\ref{Appendix:B}.

\subsubsection{Electroweak Symmetry Breaking:}
\label{sec:ewsb}
In pure supersoft supersymmetry, the operator in Eq.~\eqref{eq:dirac_gaugino} responsible for the $SU(2)$ and $U(1)$ gaugino masses also cancels their $D$-term contributions to the scalar potential, making the Higgs massless at tree level. In GSS, the Majorana adjoint masses introduced via Eq.~\eqref{eq:non_stand} disrupts the $D$-term cancellation, and the Higgs quartic coupling is non-zero. While smaller than the conventional MSSM quartic (which is already too small for the observed Higgs mass), the fact that the GSS quartic is non-zero means we do not need to rely as much on radiative stop contributions to achieve $m_h = 125\, \text{GeV}$. In order to find the conditions for electroweak symmetry breaking, we look at the quartic terms of the real scalar potential.
\begin{eqnarray}
V_1 &=&\frac{1}{32}\left[\frac{g^{\prime 2} M_{\Sigma_1}^2}{M_{\Sigma_1}^2+4 M_{D_1}^2}+\frac{g^2 M_{\Sigma_2}^2}{M_{\Sigma_2}^2+4 M_{D_2}^2}\right]\left(h_d^2-h_u^2\right)^2,
\label{eq:quartic}
\end{eqnarray}
where we have integrated out the scalar adjoint fields. In the limit $M_{\Sigma}\gg M_{D}$, we get back the MSSM quartic term $V=\frac{\left(g^2+g^{\prime 2}\right)}{32}\left(h_d^2-h_u^2\right)^2$, while the opposite limit $M_{D}\gg M_{\Sigma}$ takes us to the purely supersoft case. 

 In general, the quartic terms stabilize the scalar potential for large values of $h_u$ and $h_d$. The common wisdom in generic MSSM-like scenarios is to look for $D$-flat directions in the field space with $\left|h_u\right|=\left|h_d\right|$, where the quartic contributions vanish. The quadratic pieces need to be positive along the $D$-flat directions in order for the potential to be bounded from below. Also, for EWSB, one linear combinations of $h_u$ and $h_d$ must have a negative squared near $h_u=h_d=0$. As the quadratic part of the GSS Higgs potential is identical to the MSSM,  the stabilization and EWSB conditions~\cite{Martin:1997ns} are as well, namely:
\begin{eqnarray}
b_{\mu}^2 > \left(\tilde m^2_{h_u}+\mu^2\right) \left(\tilde m^2_{h_d}+\mu^2\right),
\label{eq:bmuset}
\end{eqnarray}
where all terms are evaluated at the IR scale. In practice, we will use Eq.~\eqref{eq:bmuset} to establish the allowed range of IR $b_\mu$ values that are consistent with given (UV) inputs of $\tilde m_{h_u}, \tilde m_{h_d}, \mu$ (equivalently, $\mu^0_u, \mu^0_d$) and $\tan\beta$; the $b_\mu(\MuIR)$ values can then be translated into $b_\mu(\MuInt)$ via Eq.~\eqref{eq:b2a_mu}.\\

Inspecting some of the results of this section, one may wonder how natural relations such as Eq.~\eqref{eq:logsize} actually are. While this is a legitimate question, it is not straight forward. Our goal is to map out  the viable parameter space of generalized supersoft theories. To concretely answer how tuned a particular parameter choice is, we need to know the UV theory behind the supersymmetry breaking spurions -- something beyond the scope of this paper.
\section{Numerical results}
\label{sec:5}
Having outlined the features of viable generalized supersoft spectra and identified what experimental constrains will shape the allowed parameter space, we now turn to numerics. For our numerical evaluation, we take $\MuIR\sim 1~\text{TeV}$ and $\MuInt = 10^{11}$~GeV, though in some cases we will vary the UV scale to show features of the running. The gauge couplings and SM fermion masses are also input parameters, with values~\cite{Xing:2007fb}: 
\begin{eqnarray}
&&g_3(\MuIR) = 1.21\;,\hspace{0.5cm} m_t(\MuIR) = 150.7~\text{GeV}\;, \nonumber \\
&&g_2(\MuIR) = 0.64\;, \hspace{0.5cm} m_b(\MuIR) = 2.43~\text{GeV}\;, \nonumber \\
&&g_1(\MuIR) = 0.45\;, \hspace{0.5cm} m_{\tau}(\MuIR) = 1.77~\text{GeV}\;.
\label{eq:incoupl}
\end{eqnarray}
We define the Yukawa couplings as $y_{u/d}=m/v_{u,d}$, where 
\begin{eqnarray}
&&v_u= \frac{v\sin\beta}{\sqrt{2}}\;,\hspace{0.5cm} v_d= \frac{v\cos\beta}{\sqrt{2}}\;, 
\end{eqnarray}
with $v=246~\text{GeV}$. This leaves us with 9 additional inputs. 
\begin{align}
\mu^0_u,\, \mu^0_d,\,  M^0_{D_a},\, M^0_{\Sigma_a},\, \tan\beta,
\end{align}
subject to the requirement that $|\mu^0_d|  > |\mu^0_u|$.

\subsection{Right-handed slepton masses}

Turning first to the right handed sleptons, we show the running of the right handed slepton mass-squared $\left(\text{scaled by}~\mathcal{S}_0\right)$ in the left panel of Fig.~\ref{sec:first_sec_sfermions} as a function of the UV scale and assuming three different values of $\tan\beta$: 2.5 (red-solid), 20 (blue-dashed), 40 (black-dotted). From these curves, one can readily estimate $\mathcal{S}_0$ in order to obtain right handed slepton mass around 100 GeV. 
\begin{figure}[h]
\centering
\includegraphics[width=\textwidth]{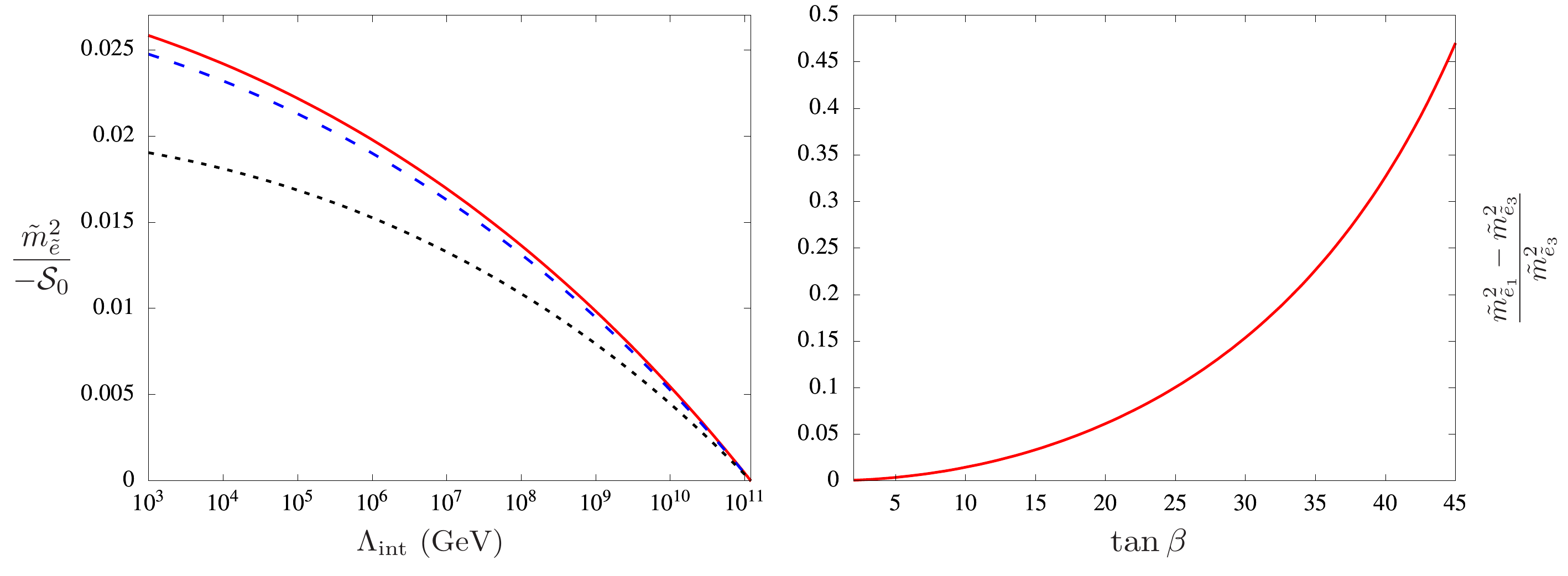}
\caption{In the left panel we show the running of the right handed slepton mass square scaled by $\mathcal{S}_0$ as a function of the intermediate scale for different $\tan\beta$ values of 2.5 (red-solid), 20 (blue-dashed) and 40 (black-dotted). In the left panel we present the mass splitting between the first and third generation of right handed sleptons as a function of $\tan\beta$. The $\tan\beta$ dependence manifests itself in the RGEs.}
\label{fig:slepton_higgsino}
\end{figure}

The solution of the full RGE has two effects not present in Sec.~\ref{sec:first_sec_sfermions} : i.) the running of $\mathcal S$, and thereby the running of all soft masses, inherits a $\tan\beta$ dependence due to the Yukawa couplings in Eq.~\eqref{eq:b2a_sr}, ii.) the running of the third generation sleptons explicitly depends on the Yukawa couplings, while the first and second generations do not. This dependence is $\tan\beta$-dependent and breaks the mass degeneracy among slepton generations, as exhibited in the right panel of Fig.~\ref{fig:slepton_higgsino}. A mass splitting between sleptons has the potential to be phenomenologically dangerous in generalized supersoft scenarios,  where -- as discussed earlier -- the largest regions parameter space will have the bino (LSP) and lightest right handed selectron nearly degenerate. As such, a mass splitting between slepton generations means the heavier sleptons are slightly heavier than the LSP and can decay $\tilde\ell\rightarrow\ell\tilde\chi_1^0$. For sleptons in the $100$ GeV range, bounds from lepton plus missing energy searches are quite stringent~\cite{Heister:2001nk,Aad:2015eda}. To avoid these bounds without raising the overall mass scale, we need to quench the splitting by restricting parameters to low to moderate $\tan\beta$. As we shall see in the next section, the small $\tan\beta$ region also well motivated from the perspective of Higgs mass.

 To obtain a rough estimate on the size of $\mathcal{S}_0$, we present two cases with different values of $\tan\beta$: 2.5 and 40. To obtain a right slepton $\MuIR$ mass of around 100 GeV for these $\tan\beta$ values, one would require
\begin{equation}
\sqrt{\left|\mathcal{S}_0\right|}=\left(\frac{\tilde m_{\tilde e}}{100\,\text{GeV}}\right) \times 
   \begin{cases}
     \frac{100}{\sqrt{0.026}}
    & \sim 620\,\text{GeV} \,\text{for}\,\tan\beta=2.5
    \\[5pt]
        \frac{100}{\sqrt{0.019}}
    & \sim 725\,\text{GeV} \,\text{for}\,\tan\beta=40 
    \end{cases}
    \label{eq:sizeofs}
\end{equation}
For $\tan\beta=2.5$, the effect of the tau Yukawa coupling in the RGEs can be neglected to a very good approximation. Hence, typical values such as $\mu_d^0=800$ GeV and $\mu_u^0=500$ GeV satisfies the left hand side of eq. (\ref{eq:sizeofs}). Furthermore, requiring the right-sleptons to be nearly degenerate results in
\begin{equation}
\frac{\left|\Delta \tilde m_{\tilde e}\right|^2}{\tilde m_{\tilde e}^2} < 1\% \implies \tan\beta \lesssim 10.
\label{eq:degen}
\end{equation}
From Eq.~(\ref{eq:degen}) and for $\tilde m_{\tilde e}\sim 100$~GeV, the degeneracy between the slepton generations turn out to be around 10 GeV. In the next section we look at the DM constraints which gives a range of the LSP-right slepton masses where relic density can be satisfied.

\subsection{Dark Matter constraints}
In the absence of coannihilation, the relic density of a bino-like neutralino can be approximated as~\cite{ArkaniHamed:2006mb}
\begin{eqnarray}
\Omega_{\tilde\chi_1^0}h^2 \simeq \frac{2.16\times 10^{-5} x_f^2}{\left|N_{11}\right|^4}\left(\frac{m_{\tilde f}}{100~\text{GeV}}\right)^2\frac{(1+r)^4}{r(1+r^2)},
\label{eq:relic-ann}
\end{eqnarray}
where $x_f$ represents the freeze-out epoch and $r=m^2_{\tilde\chi_1^0}/\tilde m^2_{\tilde e}$. For a bino well-separated from other states, say $r < 0.8$, the present value of the observed relic density 0.1199~\cite{Ade:2015xua} readily limits $\tilde m_{\tilde e}\lesssim 100$ GeV. Such slepton masses are very tightly constrained from the LEP data~\cite{LEP:2}. As a result, in most of the allowed parameter space of slepton masses, the bino suffers from overabundance. 

In our setup, the bino is nearly mass-degenerate with right handed sleptons, and consequently coannihilation becomes important. Roughly,  whenever $\delta m\equiv \tilde m_{\tilde e}-m_{\tilde\chi_1^0}\sim T_f$, then these slightly heavier degrees of freedom are thermally accessible and are therefore nearly abundant as the relic species. Since $T_f\sim m_{\tilde\chi_1^0}/25$~\cite{ArkaniHamed:2006mb}, we find the degree of degeneracy needed for coannihilation is 

\begin{eqnarray}
\frac{\tilde m_{\tilde e}-m_{\tilde\chi_1^0}}{m_{\tilde\chi_1^0}}&=&
\frac{\delta m}{m_{\tilde\chi_1^0}}=\frac{T_f}{m_{\tilde\chi_1^0}}
\sim 5 \%.
\end{eqnarray}
Once coannihilation becomes important, the slepton self interactions and interactions with the bino~\cite{Ellis:1998kh} set the relic abundance. These additional interactions relax the bound on the slepton masses and overabundance issue can be avoided if
\begin{eqnarray}
\tilde m_{\tilde e}\sim m_{\widetilde\chi_1^0}\lesssim 500 \textrm{GeV},
\end{eqnarray}
which is well above the current LHC bound on sleptons nearly degenerate with the LSP~\cite{Aad:2015baa}. For a dedicated and more robust analysis, we implemented the effective $\MuIR$ model in {\tt SARAH-4.11.0}~\cite{Staub:2008uz,Staub:2012pb} and generated the spectrum using {\tt SPheno-3.3.3}~\cite{Porod:2003um,Porod:2011nf}. We have varied the masses of the lightest neutralino state and the slepton masses (assumed to be degenerate). If the LSP is neutralino then only the spectrum file is fed to {\texttt{micrOMEGAs-v3}}~\cite{Belanger:2001fz} for computing relic abundance and dark matter direct detection rates. We find that coannihilation works efficiently for $m_{\tilde\chi_1^0}\sim m_{\tilde e}\sim 400$ GeV. This immediately sets $\xi_d (\MuInt)\sim 2$~TeV and $\mu(\MuIR)\sim 600$~GeV. This limit is obtained by considering again $\xi_u(\MuInt)\sim 0$.

\subsection{Higgsino masses}
The Higgsino mass is essentially tied with the right handed slepton masses through their initial conditions. 
\begin{figure}[h]
\centering
\includegraphics[width=1.\textwidth]{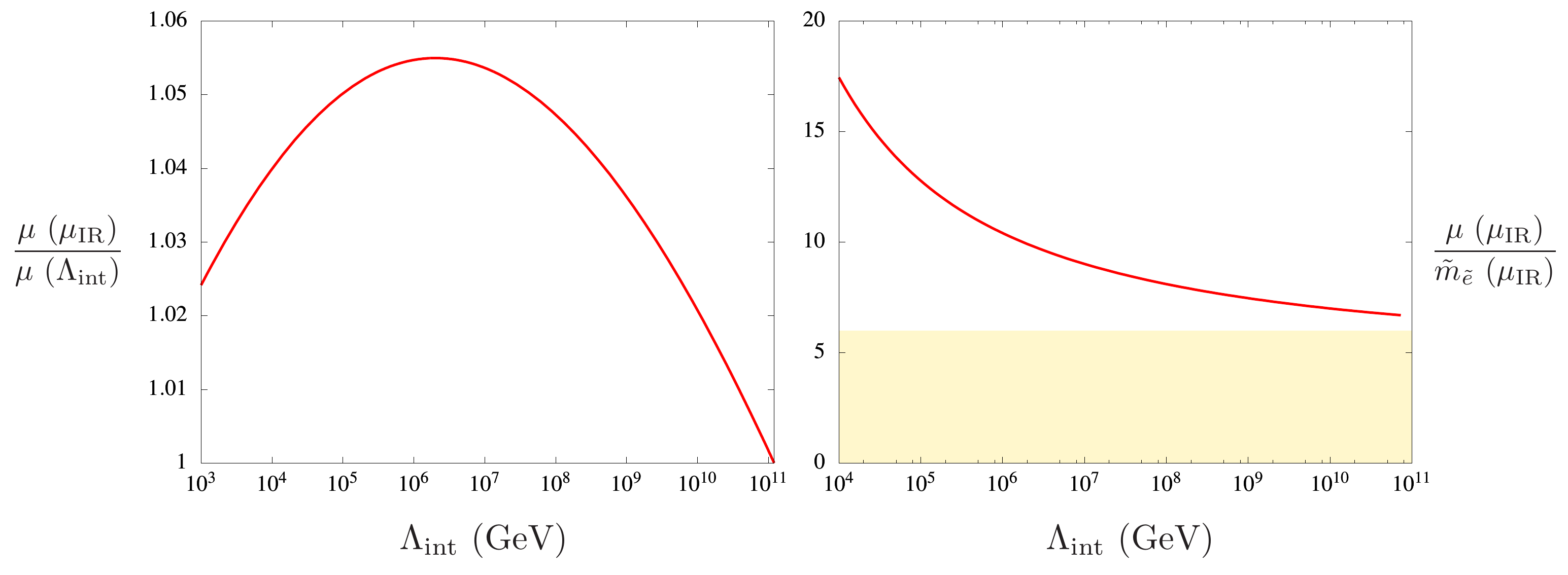}
\caption{In the left panel of the figure we show the running of the higgsino mass scaled with its initial condition for a fixed value of $\tan\beta=2.5$. On the right panel we elucidate how the low energy constraints can be translated to a bound on the intermediate scale. These constraints can come from either DM direct detection results or collider experiments. The yellow shaded region is excluded from the direct collider searches for right slepton/LSP masses close to 100 GeV for the same initial conditions as discussed before.}
\label{fig:higgsino}
\end{figure}
Assuming again $\mu_u^0=500$ GeV and $\mu_d^0=800$ GeV, the right panel of the Fig.~\ref{fig:higgsino} shows that the Higgsino mass at $\MuIR$ is driven by
\begin{equation}
\mu(\MuIR)\Big|_{\tan\beta=2.5}\sim 1.025\times \frac{\mu_d^0+\mu_u^0}{2}\sim 660~\text{GeV}.
\end{equation}
Another way to make higgsinos heavy would rely on the modification of the messenger scale. 

In the right panel of Fig.~\ref{fig:higgsino} we show how low energy constraints can significantly constrain the messenger scale in our scenario. The constraints are two folds, first from direct detection experiments. The limit from direct detection experiment constrains the higgsino fraction in the lightest neutralino. As a result, higgsino should be heavier compared to the LSP or the lightest slepton. In our case, the lightest neutralino is a predominant mixture of the bino and singlino gauge eigenstates and the higgsino admixture is negligible. Therefore, the stringent constraint from the DM direct detection can be avoided easily. However,  to push the Higgs mass to the observed value, as elaborated in the next section, we require introducing new superpotential terms. For larger couplings, this increases the higgsino component in the lightest neutralino state. Secondly, collider experiment can also provide stringent constraint on the ratio of the higgsino and slepton masses. For example, our spectrum has the following hierarchical structure where NSLPs are the right handed slepton and LSP is the bino-singlino admixture neutralino. Higgsinos are heavier than the sleptons. Such a higgsino after electroweak production can decay to a $Z\tilde\chi_1^0$ or $h\tilde\chi_1^0$~\cite{Aad:2014vma,Aad:2015jqa}. However, all these modes are phase space suppressed for $\tilde\chi_2^0\sim\tilde\chi_3^0\sim 200$~GeV. In such cases the dominant decay mode would be $\tilde\ell\ell$. The limits on this particular final state is very robust~\cite{Aad:2015eda}. In our case, we took a conservative approach and used $\mu(\MuIR)/\tilde m_{\tilde\ell_{R}}(\MuIR)\sim 6$. For our choice of parameters, we observe from Fig.~\ref{fig:higgsino} that the UV scale should be less than $10^{11}$ GeV or so. Changing the initial values would modify the results as the dependence of these two parameters are different for higgsino and slepton mass runnings.

\subsection{Dark matter direct detection}
Our framework also needs to be consistent with the null results in DM direct detection. When the squarks are heavy, the spin-independent interaction between DM and nuclei comes from Higgs exchange and thus it depends crucially on the Higgs coupling to the lightest neutralino.  The Higgs-neutralino coupling, in turn, depends on the higgsino mass parameter. For a given LSP bino mass we can translate limits from direct detection directly into limits on the higgsino mass. The spin-independent cross-section can be well approximated by the following relation~\cite{Jungman:1995df}
\begin{eqnarray}
\sigma_{\text{SI}}&\simeq&\frac{8 G_F^2}{\pi}M_Z^2 m^2_{\text{red}}\frac{F_h^2 I_h^2}{m_h^4},
\end{eqnarray}
where $G_F, M_Z, m_{\text{red}}$ are the Fermi coupling constant, $Z$ boson mass, and DM-nucleon reduced mass, and $F_h$, $I_h$ are coupling and kinematic factors.  In the limit when the wino and heavy Higgs are heavy and effectively decoupled (in addition to the squarks), 
\begin{equation}
F_h = - N_{11} N_{14} \sin\theta_W\;, \hspace{0.5cm}
I_h = \sum_{q} k_q^h m_q \langle N|\bar{q}q|N\rangle.
\end{equation}
Here, $N_{11}, N_{14}$ are the bino and higgsino fraction in the lightest neutralino, $k^h_{\text{up-type}}=\cos\alpha/\sin\beta$, $k^h_{\text{down-type}}=-\sin\alpha/\cos\beta$, and we have already assumed $\cos\alpha\rightarrow 1$ and $\sin\alpha\rightarrow 0$ by decoupling the heavy Higgs. Because of the Dirac nature of the gaugino masses and the extra interactions involving the adjoint (e.g., $SH_uH_d$ coupling in the superpotential), the neutralino mass matrix no longer have the MSSM form. Taking the ratio of the DM direct detection in GSS to the prototypical MSSM, the only factors that don't drop out are the $N_{11}, N_{14}$,
\begin{eqnarray}
\frac{\sigma^{\text{GSS}}_{\text{SI}}}{\sigma^{\text{MSSM}}_{\text{SI}}}\simeq \frac{\left| N^{GSS}_{11}\, N^{GSS}_{14}\right|^2}{\left| N_{11} N_{14}\right|^2}.
\end{eqnarray}
For typical values $\mu=300$~GeV, $M_{\tilde\chi_1^0}\sim 150$~GeV, $\tan\beta=4$, the direct detection cross-section in GSS turns out to be an order of magnitude less than in  the MSSM, as the bino-singlino mixing in GSS dominates over bino-higgsino mixing. This implies lower higgsino masses are possible in our framework.  

The constraints on the right handed slepton masses and the higgsino mass can be effectively translated to constrain the two input parameters $\mu_u^0$ and $\mu_d^0$. 
\begin{figure}[h]
\centering
\includegraphics[width=0.65\textwidth]{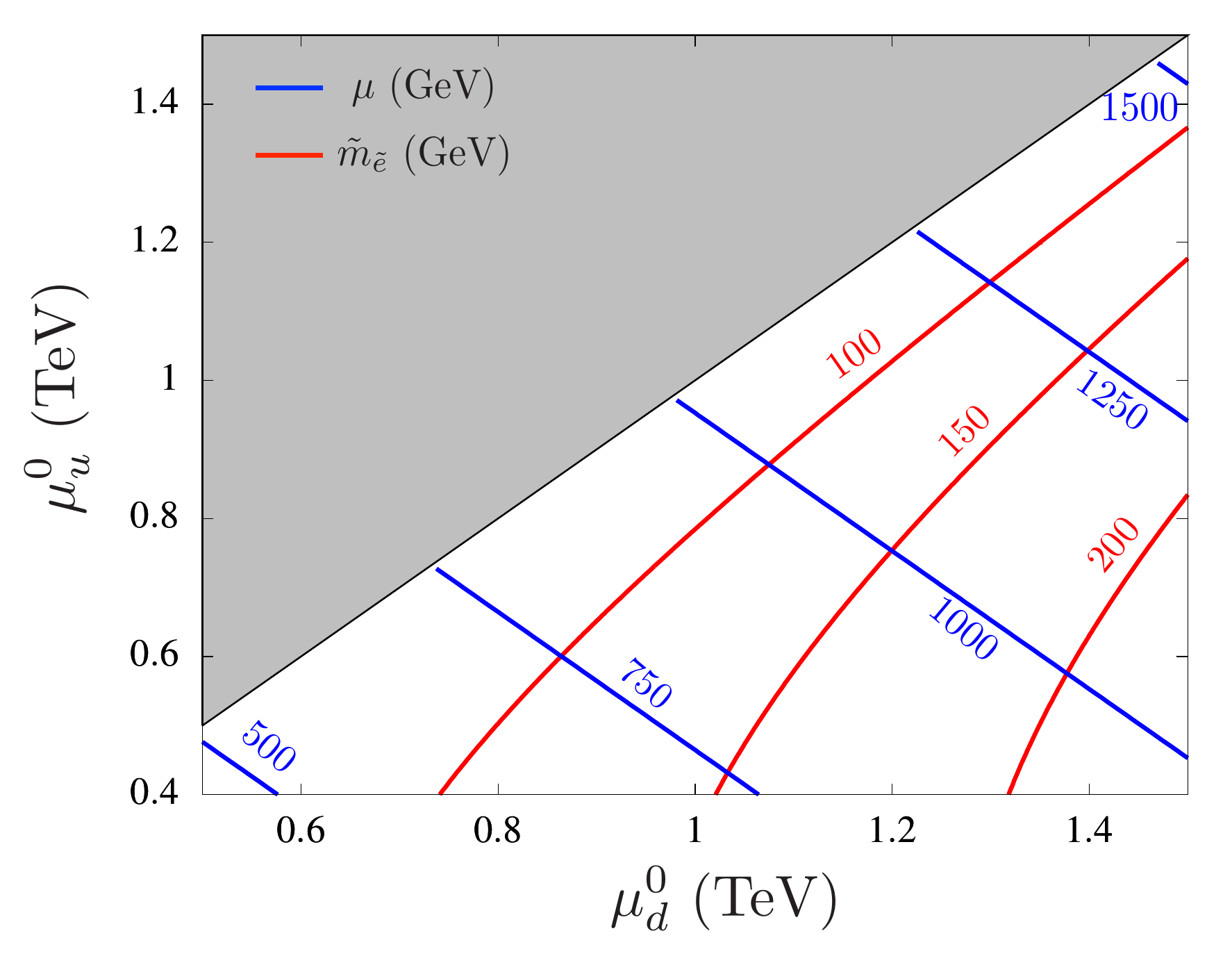}
\caption{We show  the contours of right handed slepton mass (red) and higgsino mass (blue) in the $\mu_u^0$-$\mu_d^0$ plane.  To obtain positive definite mass for the right handed sleptons one needs to have $\mu_d^0>\mu_u^0$, therefore, ruling out the shaded region. For definite values of right handed slepton and higgsino masses, designated by the crossing of the contours, the two input parameters $\mu_u^0$ and $\mu_d^0$ can be readily obtained. We have fixed $\Lambda_{\text{int}}=10^{11}~\text{GeV}$ and $\tan\beta=2.5$. }
\label{fig:third}
\end{figure}
In Fig.~\ref{fig:third} we show the contours of right handed slepton mass (red) and higgsino mass (blue) in the $\mu_u^0$-$\mu_d^0$ plane. As already stated, one needs to have $\mu_d^0>\mu_u^0$ in order to get positive definite right handed slepton mass. As a result, the grey shaded region is not viable. The masses of right handed sleptons and the LSP should be nearly degenerate in order to satisfy the relic abundance. Moreover, DM direct detection experiments sets a limit on the bino-higgsino mixing and effectively on the higgsino mass parameter, $\mu$. Therefore, given the values of $m_{\tilde\ell_R}$ and $\mu$ one can readily constrain the input parameters $\mu_u^0$ and $\mu_d^0$.  

\subsection{Higgs mass and new superpotential terms}
\label{sec:mH}
In previous sections we have shown how the inputs $\mu^0_u, \mu^0_d$, the Dirac and Majorana gaugino masses and, to some extent, $\tan\beta$ are constrained by collider physics and Dark Matter. What remains to be done is to see what Higgs masses are possible in the `surviving' regions. As already mentioned in sec.(\ref{sec:ewsb}), the traditional MSSM Higgs quartic terms get depleted due to the presence of Dirac gauginos. However, new superpotential couplings such as $\lambda_S$ generates additional quartic terms which are NMSSM like. Under some simplified assumptions such as i.) integrating out the adjoint scalar fields, ii.) assuming $\lambda_T=0$, for simplicity, the Higgs mass can be obtained by diagonalizing the scalar mass matrix in the basis $(h_u,h_d)$ given in Appendix~\ref{Appendix:C}.

To fully answer the question of the Higgs mass, we need to go beyond tree-level. The largest loop-level contribution comes from the top squarks and is governed by their overall mass and degree of $\widetilde t_L-\widetilde t_R$ mixing. In GSS, the top squark mass matrix, neglecting $D$-term contributions, is:
\begin{equation}
\renewcommand{\arraystretch}{1.2}
m^2_{\tilde t} = \left(\begin{array}{cc}
m_{Q_3}^2+m_t^2 & -\tilde\xi_u y_t v_d \\
-\tilde\xi_u y_t v_d & m_{\bar u_3}^2+m_t^2\\
\end{array}\right),
\end{equation}
where the terms $\tilde\xi_u$ is the supersymmetry breaking trilinear interaction originating from Eq.~\eqref{eq:GSSLagrangian}. It is well known that the top squark contribution to the Higgs mass is largest when there is substantial  $\widetilde t_L-\widetilde t_R$ mixing~\cite{Martin:1997ns}. In GSS, the mixing angle in the stop-sector,
\begin{equation}
\tan 2\theta=\frac{-2\tilde\xi_u y_t v_d}{m_{Q_3}^2-m_{\bar u_3}^2},
\end{equation}
is proportional to $v_d \sim \cos{\beta}$, while the mixing angle in the MSSM $\sim \sin\beta$. The difference can be traced to the unusual structure of the scalar trilinears in GSS and, since one usually wants to take $\tan\beta$ large to maximize the tree level Higgs quartic, leads to suppressed stop sector mixing. Suppressed stop sector mixing can be overcome by taking large $\tilde\xi_u$, though this is an unattractive option as it will increase the traditional fine tuning measure of the setup. Moreover, even if we set aside tuning concerns for the moment, we find that even very large values of $\tilde\xi_u$ are unable to push the Higgs mass to more than 100 GeV (recent works, studying the phenomenological implications of  such non-standard soft supersymmetry breaking terms in MSSM can be found in~\cite{Chattopadhyay:2016ivr,Chattopadhyay:2017qvh}). In addition to the tree level terms, we have also included two loop corrections from the stop sector~\cite{Heinemeyer:1999be} (for three loop corrections in MSSM see~\cite{Feng:2013ena}) and considered $m_{\tilde t_1}\sim m_{\tilde t_2}\sim 1.8$~TeV.\footnote{In the presence of additional superpotential interactions (Eq.~\eqref{eq:theWnew}), the electroweak adjoints also contribute to the Higgs mass and can in principle be included. At one loop, these contributions are $\propto \lambda^4_S, \lambda^4_T$ and are logarithmically sensitive to the difference between the adjoint fermion and scalar masses. However, unlike in the top-stop sector, the ratio of adjoint fermion to scalar masses is $\mathcal O(1)$, thereby suppressing the adjoint contribution to the Higgs mass. We have therefore neglected this (positive definite) piece for simplicity. } Higher Higgs masses are possible if we consider heavier stops, though at the expense of increased tuning.

Hence, the most natural way to increase the Higgs mass is to extend the theory with additional $F$-terms, as in the NMSSM. In GSS, this extension requires no new matter as the theory already contains a $SU(2)$ singlet and triplet superfield. Including interactions between these superfield and the Higgses, the GSS superpotential is modified to
\begin{eqnarray}
W_{GSS}=W + \lambda_S  S\, H_u\, H_d + \lambda_T H_u\,T\,H_d,
\label{eq:theWnew}
\end{eqnarray}
with $W$ given in Eq.~\eqref{eq:GSSLagrangian}. The modified superpotential generates a new tree level quartic for the Higgs. Specifically, taking $\lambda_S \ne 0, \lambda_T = 0$ for simplicity, two new Higgs potential terms are generated, 
\begin{align}
V_2 &= \frac{\left|\lambda_S\right|^2}{4}\left[\frac{4 M_{D_1}^2}{M_{\Sigma_1}^2+4 M_{D_1}^2} h_u^2 h_d^2+\frac{\mu M_{\Sigma_1}}{M_{\Sigma_1}^2+4 M_{D_1}^2}h_u h_d \left(h_u^2+h_d^2\right)-\frac{\mu^2}{M_{\Sigma_1}^2+4 M_{D_1}^2}\left(h_u^2+h_d^2\right)^2\right], \nonumber \\
V_3 &= -\frac{\sqrt{2}g^{\prime}\lambda_S}{4}\frac{M_{D_1}}{M_{\Sigma_1}^2+4 M_{D_1}^2}\left[\mu\left(h_u^2+h_d^2\right)-M_{\Sigma_1}h_u h_d\right]\left(h_d^2-h_u^2\right).
\end{align}
In the limit $M_{D}\gg M_{\Sigma}, \mu$, the quartic no longer vanishes but instead takes on a NMSSM form,  e.g., $V=\frac{\left|\lambda_s\right|^2}{4} h_u^2 h_d^2$\, and can generate a large tree level Higgs mass.\footnote{The $\lambda$-dependent pieces do not vanish in the D-flat limit so they will alter the stabilization conditions, as well as the relation between the $b_{\mu}$ and the other inputs.}

\begin{figure}[h]
\centering
\includegraphics[width=\textwidth]{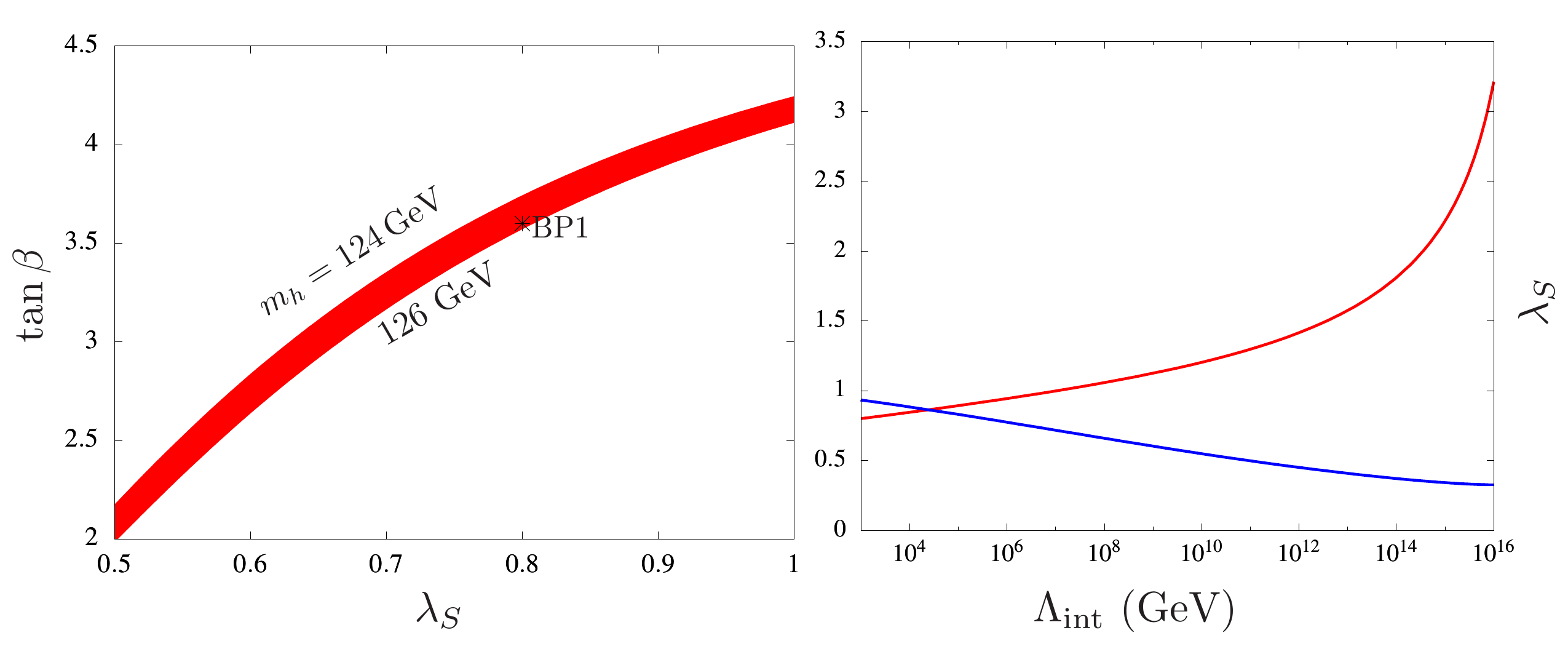}
\caption{Left: Higgs mass with new tree level quartic contributions after taking into account two loop corrections from the stop sector ($\tilde m_{\tilde t}\sim 1.8~\text{TeV}$). The MSSM tree level contribution is depleted due to our choice of $M_\Sigma$ and $M_D$'s (same as BP1). However, additional $F$-term contributions occur from the superpotential coupling $\lambda_S$. Right: Running of superpotential coupling $\lambda_S$ (shown by the red-solid curve on the left panel) and top Yukawa coupling (blue-solid curve). We have fixed $\lambda(\MuIR)=0.8$ and $\tan\beta=3.6$ (same as BP1) to satisfy Higgs mass constraints. }
\label{fig:higgsmass}
\end{figure}
In the left panel of Fig.~(\ref{fig:higgsmass}) we show the Higgs mass with new tree level quartic contributions after taking into account two loop corrections from the stop sector ($\tilde m_{\tilde t}\sim 1.8~\text{TeV}$). The MSSM tree level contribution gets reduced due to the presence of $M_{\Sigma_i}$ and $M_{D_i}$, however additional $F$-term contributions from the superpotential coupling $\lambda_S$ helps.  The values shown in Fig.~\ref{fig:higgsmass} are the IR values, so one might worry that the relatively large $\lambda$ values needed for $m_h = 125\, \text{GeV}$ grow even larger in the UV and lead to a violation of perturbativity.  In addition to their own running, the $\lambda$ couplings modify the anomalous dimensions of the Higgs fields and thus contribute to the running of the top and bottom Yukawa couplings:
\begin{eqnarray}
\beta\left[\lambda_S\right] &=& \lambda_S \left[3 y_t^2+3 y_b^2-3g^2-\frac{3}{5}g_1^2+4\lambda_S^2+6\lambda_T^2\right], \nonumber \\
\beta\left[\lambda_T\right] &=& \lambda_T \left[3 y_t^2+3 y_b^2-7g^2-\frac{3}{5}g_1^2+2\lambda_S^2+8\lambda_T^2\right], \nonumber \\
\beta\left[y_t\right] &=& y_t \left[6 y_t^2+ y_b^2-\frac{16}{3}g_3^2-3g^2-\frac{13}{15}g_1^2+\lambda_S^2
+3\lambda_T^2\right] \;. 
\label{eq:running_lam}
\end{eqnarray}
We reiterate, the RGEs of the scalar masses are shown to run from $\MuInt\sim 10^{11}$ GeV, therefore integrating out any hidden sector effects. However, $\lambda$'s are superpotential coupling and should remain perturbative upto the GUT scale. The running of the singlet coupling to the Higgs fields is shown in Eq.~(\ref{eq:running_lam}) and depicted in the right panel of Fig.~\ref{fig:higgsmass}. We have chosen $\lambda(\MuIR)=0.8$ and $\tan\beta=3.6$ (same as BP1) to satisfy Higgs mass constraints. The coupling remains perturbative up to the GUT scale. However, due to larger group theory factors in the anomalous dimension the triplet coupling diverges more rapidly, becoming nonperturbative at an energy close to $10^{10}$~GeV unless additional structure is added to the theory.

The new superpotential couplings generate new trilinear interactions in the scalar potential and modify the running of the Higgs soft masses $\tilde m_{h_u}$, $\tilde m_{h_d}$  and couplings $\tilde\xi_u, \tilde\xi_d$. The rest remains unaffected.
The additional couplings in the Lagrangian 
\begin{equation}
\mathcal L \rightarrow \mathcal L -\lambda_S \xi_u \left| h_u \right|^2 S - \lambda_S \xi_d \left| H_d\right|^2 +\text{triplet trilinear terms}
\end{equation}
The modified RGE are presented in Appendix~\ref{appendix:D}. One important thing to note that the presence of the singlet field and non-standard soft terms can give rise to dangerous tadpole diagrams which might destabilize the hierarchy~\cite{Hetherington:2001bk,Dedes:2000jp}. We discuss such issues in Appendix~\ref{appendix:E}.

\subsection{Benchmark Points}
\label{sec:bench}

Finally, we provide two benchmark points to show that the framework is consistent with all the phenomenological observations including the Higgs mass, DM relic density, DM direct detection and collider results. The UV and IR parameters for these benchmarks, along with $m_h, \Omega\, h^2$ and the DM-nucleon spin-independent cross section are shown in Table~\ref{tab:bench}. As discussed in Sec.~\ref{sec:mH}, a viable Higgs mass is most naturally reached in GSS when one admits superpotential interactions between the $SU(2),  U(1)$ adjoint superfield and the Higgses. While either $\lambda_S$, $\lambda_T$ or both can be used, however, $\lambda_T = 0$ is the safer choice, in the sense that it requires fewer assumptions about the UV. Therefore, both of the benchmark setups have $\lambda_S \ne 0, \lambda_T = 0$.

\begin{table}[t!]
\centering
\begin{tabular}{ |c|c|c|c| } 
\hline
Parameters ($\MuInt$) & BP1 & BP2 \\
\hline
$\MuInt$ & $10^{11}$ GeV & $10^{11}$ GeV \\
$M_{D_{3}}$ & $2.48$ TeV & 2.65 TeV \\ 
$M_{D_{2}}$ & $4.72$ TeV & 5.20 TeV \\ 
$M_{D_1}$ & 556 GeV & 567 GeV \\
$M_{\Sigma_{3}}$ & 0 & 940 GeV \\
$M_{\Sigma_{2}}$ & 0 & 3.45 TeV \\
$M_{\Sigma_1}$ & 1.97 TeV & 1.30 TeV \\
$\tilde m_{\tilde q,\tilde u, \tilde d, \tilde \ell, \tilde e}$ & 0 & 0 \\ 
$\mu_d^0$ & 1.35 TeV & 1.43 TeV \\
$\mu_u^0$ & 1.15 GeV & 1.0 TeV \\
$b_\mu$ & $0.739^2~\text{TeV}^2$  &  $0.858^2~\text{TeV}^2$\\
\hline
Parameters ($\MuIR$) & BP1 & BP2\\
\hline
$M_{D_{3}}$ & $7.00$ TeV & 7.50 TeV \\ 
$M_{D_{2}}$ & $5.00$ TeV & 5.50 TeV \\ 
$M_{D_1}$ & 515 GeV & 525 GeV \\
$M_{\Sigma_{3}}$ & 0 & 7.50 TeV \\
$M_{\Sigma_{2}}$ & 0 & 5.50 TeV \\
$M_{\Sigma_1}$ & 2.50 TeV & 1.65 TeV \\
$\mu$ & 1.01 TeV & 1.056 TeV\\
$b_\mu$ & $0.749^2~\text{TeV}^2$  &  $0.867^2~\text{TeV}^2$\\
$\lambda_S$ & 0.80 & 0.7 \\
$\tan\beta$ & 3.6 & 2.9 \\
\hline
Outputs & BP1 & BP2 \\
\hline
$\tilde m_{\tilde q, \tilde u, \tilde d}$ & $1.845,1.842,1.845$ TeV & $1.823,1.817,1.824$ TeV \\
$\tilde m_{\tilde t_{1,2}}$ & $1.847, 1.85$ TeV & $1.82, 1.83$ TeV \\
$\tilde m_{\tilde\chi_{2,3}}$ & 1.016, 1.025 TeV & 1.051, 1.061 TeV \\
$\tilde m_{\tilde\ell}$ & 513.5 GeV & 514.4 GeV \\
$\tilde m_{\tilde e}$ & 110 GeV & 160 GeV \\
$\tilde m_{\tilde\chi_1^0}$ & 100 GeV & 150 GeV \\
$m_h$ & 125.6 GeV & 125.3 GeV \\
$m_H$ & 1.47 TeV & 1.55 TeV \\
$\Omega h^2$ & 0.081& 0.107\\
$\sigma_{\text{SI}}$ & $1.615\times 10^{-47}~\text{cm}^2$ & $6.17\times 10^{-47}~\text{cm}^2$ \\
\hline 
Branching Ratios & BP1 & BP2 \\
\hline
$\tilde\chi_2^0\rightarrow Z\tilde\chi_1^0$ & 72.7\% & 87.2\% \\
$\tilde\chi_2^0\rightarrow h\tilde\chi_1^0$ & 23.8\% & 10.6\%\\
$\tilde\chi_3^0\rightarrow Z\tilde\chi_1^0$ & 24.4\% & 85.0\% \\
$\tilde\chi_3^0\rightarrow h\tilde\chi_1^0$ & 70.3\% & 10.4\%\\
$\tilde\chi_1^+\rightarrow W\tilde\chi_1^0$ & 96.7\%& 96.7\% \\
\hline
\end{tabular}
\caption{Benchmark points for GSS scenarios with the two different values of $\lambda_S$. Points satisfy the Higgs mass and DM relic density and direct detection constraints. The DM direct detection limit for a 100 GeV WIMP is around $9\times 10^{-47}~\text{cm}^2$~\cite{Akerib:2016vxi}.}
\label{tab:bench}
\end{table}

Inspecting the benchmarks, both points share the feature that $\tilde q, \tilde{u},\tilde{d}, \tilde{\ell}, \tilde e$ are massless at $\MuInt$ and have $\mu^0_{d,u}$ split in a way that yields positive RH slepton masses, $\mu^0_d - \mu^0_u \sim 100\,\text{GeV}$. To augment the Higgs mass, both benchmarks have $\lambda_S \ne 0, \mathcal O(1)$. Order one values of $\lambda_S$ enhance the higgsino fraction of the lightest bino sector field~\cite{Belanger:2009wf}. Hence, to overcome stringent constraints from the DM direct detection\footnote{We have used {\texttt{micrOMEGAs-v3}} for DM direct detection cross-section. For $\sim$ 100 GeV neutralino, the cross-section should be less than $9\times 10^{-47}~\text{cm}^2$~\cite{Akerib:2016vxi}.}, the higgsino mass -- set by $\mu^0_{u,d}$ -- needs to be around a TeV.  We note in passing that such a problem does not arise when one makes the triplet coupling $\lambda_T$ large to fix the Higgs mass. As this coupling only increases the higgsino fraction in the wino-tripletino like neutralino, considered to be heavy in our case.  For both benchmarks, the $b_\mu(\MuInt)$ values have been determined numerically following the logic established in Sec.~\ref{sec:ewsb} with refinements due to $\lambda_S \ne 0$ and are similar in size to the other UV inputs. 

The largest difference between the points is the mass of the triplet and octet fermions. In BP1, $M_{\Sigma_2} = M_{\Sigma_3} = 0$ in the UV, which, since the running of these masses is multiplicative, implies $M_{\Sigma_2} = M_{\Sigma_3} = 0$ in the IR as well.\footnote{There is no phenomenological disaster associated with $M_{\Sigma_2} = M_{\Sigma_3} = 0$ as the $\Sigma_{2,3}$ fields will receive a finite, loop-level contribution from gaugino loops, see Sec.~\ref{sec:colored} and~\ref{sec:ew}} We cannot further simplify things by choosing $M_{\Sigma_1} = 0$ without destroying the see-saw mechanism in the bino sector.   A second difference between the benchmarks is the LSP mass, set to be $100\, \text{GeV}$ in BP1 and $150\,\text{GeV}$ in BP2. The fact that the LSP mass is higher in BP2 while the higgsino mass remains the same as in BP1 results in slightly enhanced direct detection limit. 

Finally, one set of bounds we have yet to address are the LHC limits on electroweakinos. The bounds on chargino/heavier neutralinos ($\tilde{\chi}^0_2, \tilde{\chi}^0_3$) production can be quite stringent, $\sim 1\, \text{TeV}$ if the chargino/neutralinos decay primarily to sleptons~\cite{Sirunyan:2017lae, ATLAS:2017uun}. Fortunately, electroweakino decays to sleptons are rare in our setup. As we have seen, collider bounds on left handed sleptons push the mass of the winos into the multi-TeV range, while DM and Higgs constraints prefer higgsinos at $1\, \text{TeV}$. As a result, $\tilde{\chi}^{\pm}, \tilde{\chi}^0_2, \tilde{\chi}^0_3$ are predominantly higgsino and thus have Yukawa-suppressed couplings to SM fermions (this is exacerbated by the small $\tan\beta$ in BP1 and BP2). With sfermion-fermion decays suppressed, the charginos and neutralinos decay preferentially to gauge bosons and Higgses, modes with much looser constrains. The dominance of the gauge/Higgs boson branching ratios of $\tilde{\chi}^{\pm}, \tilde{\chi}^0_2, \tilde{\chi}^0_3$ in BP1 and BP2 can be seen in the bottom rows of Table~\ref{tab:bench}.

These two benchmarks have been chosen with particles sitting just outside the existing bounds. In the near future, both points would be exposed through jets plus missing energy searches (sensitive to $\tilde q, \tilde u, \tilde d$) or leptons plus missing energy (sensitive to $\tilde{\ell}$). However, the squarks and left handed sleptons can easily been taken heavier without ruining the main features of GSS, namely the near degeneracy of the right handed sleptons with the LSP. Compressed spectra, especially among weakly interacting states, are hard to probe at the LHC, though studies with displaced vertices~\cite{Liu:2015bma, Bramante:2015una}, soft-tracks~\cite{Chakraborty:2016qim,Chakraborty:2017kjq}, or initial state radiation~\cite{Martin:2007gf,Schwaller:2013baa,Han:2014kaa,Han:2014xoa, Delgado:2016gqn} can be a useful tool to look for such regions. Should a compressed slepton-LSP sector be discovered at the LHC, the next step towards singling out GSS as the underlying framework would be to verify the hierarchical structure of the spectrum, for example, a right handed slepton signal without any sign of squarks, left handed sleptons, or gauginos. As the absence of other states is a rather unsatisfactory discriminator among models, a more concrete signal is the presence of $SU(2)$, $SU(3)$ adjoint scalars. The masses of these states is more model dependent (e.g. compare BP1 and BP2) and it is possible that they are light enough to yield a signal at the LHC~\cite{Plehn:2008ae}, however it is likely that a future, higher energy machine is required to explore the spectrum fully.

\section{Summary and conclusion}
\label{sec:6}
The recent run of the LHC has put stringent constraints on the superpartner masses. In fact, the strongest limit is drawn when the gluinos/squark masses are well separated from the LSP mass, reaching almost 2 TeV. A heavy gluino raises the soft mass of Higgs fields through renormalization group evolutions. As these parameters are a measure of the fine tuning, the non-observation of the superpartners has resulted into finely tuned regions of parameter space for most supersymmetry models. Frameworks with supersoft supersymmetry with Dirac gauginos are well motivated in this light. The supersoft nature of the gluinos ensure that the gaugino mediated correction to the squark masses are finite and not log enhanced. Therefore, Dirac gluinos can be naturally heavy. Consequently, the pair production of the gluinos goes down significantly due to kinematic suppression. Moreover, the production of same chirality squarks are also less as this requires chirality flipping Majorana gaugino masses in the propagator. The reduction in the production cross-section of the squarks weakens the constraints on the squark masses significantly. Hence, supersoft models are often coined as `supersafe' in the literature. An additional virtue of this framework is flavor and CP violating effects are well under control. 

However, supersoft frameworks also suffer from a few drawbacks.  First, the Guidice-Masiero mechanism is unavailable so viable $\mu$ values require a conspiracy between the supersymmetry breaking scale and the Planck scale. Further, the natural $D$-flat direction of the Higgs potential sets the tree level quartic term to zero, making it very difficult to fit the observed Higgs mass of 125 GeV. Finally, supersoft models contain additional scalars in the adjoint representation of the SM gauge group that often acquire negative squared masses, resulting in a color breaking vacuum. An interesting way to resolve these three issues is to supplement the theory with additional, potentially non-supersoft operators involving the same $D$-term vev used to generate gaugino masses, the so-called generalized supersoft framework. The additional operators in GSS generate $\mu$-terms proportional to the supersymmetry breaking vev and positive definite masses (squared) for the scalar components of the adjoint chiral superfields. While economically solving the $\mu$ and adjoint masses issues is a step forward, Dark Matter remains an issue in GSS; right handed sleptons only receive mass through the finite correction from the bino and therefore seem to be destined to be the LSP unless the mediation scale is low.

In this work we mapped out the parameter space of GSS, paying particular attention to the DM problem. We have shown that bino LSP can be avoided while maintaining a high mediation scale and without new fields in parameter regions where the two GSS $\mu$-terms are unequal. If $\mu_u\neq \mu_d$, supersoftness is lost and loop suppressed, log divergent pieces from the hypercharge $D$-term contribute to the running of the scalar masses. For the right choice of inputs, these hypercharge contributions can lift up the RH slepton mass above the bino. Focusing on the region where the bino is the LSP, we explored how well the setup can satisfy additional constraints such as the $125\, \text{GeV}$ Higgs mass, correct relic abundance -- achieved whenever the bino and RH slepton masses are similar and coannihilation becomes important --  and compatibility with the latest LHC results. 

The RH slepton masses are controlled by the difference of the UV $\mu$ parameters while the higgsino mass is set by the sum. If we insist that the LSP bino is a valid thermal relic DM that escapes all direct detection bounds, the two constraints become tightly correlated, since the slepton mass controls the degree of coannhiliation while the higgsino mass controls the strength of the Higgs-exchange DM-nucleon interaction. Within this parameter region, we find squark and slepton collider bounds can be satisfied, but the Higgs mass is generically too low unless one resorts to large loop corrections. Rather than resort to heavy stops, we showed how additional, NMSSM-type interactions can be used to lift the Higgs mass. These NMSSM-type interactions require no additional field content, as the bino's Dirac partner is a gauge singlet. The interpolation between IR constraints and UV inputs requires some care, as non-standard supersymmetry breaking interactions are generated whenever $\mu_u\neq \mu_d$ and enter non-trivially into the RGE.

Even though we supplement this work with full numerical solutions, we focus rather on calculating general features of the spectrum, which we derive using analytical solutions whenever we can after making various simplifying assumptions. In fact, most of the features of the electroweak spectra get captured even after these simplifications. Instead of scanning the full parameter space numerically for allowed regions, this approach allows us to generate intuitions about how to convert various experimental bounds into bounds in the UV parameter space within this framework. 
Before we conclude, note that while in this work we focus on charting the parameter space for viable spectra, it still remains a open question how these new $\mu$-type operators can be generated in a calculable UV theory, especially in the limit  $\mu_u\neq \mu_d$.  While it is not hard to speculate a strongly coupled theory giving rise to these operators, where the supersymmetry breaking spurion arises from real operators of hidden sector, finding a concrete model is challenging.

\section*{Acknowledgements}
The work of AM was partially supported by the National Science Foundation under Grant No. PHY-1520966. 
\appendix
\section{Neutralino and Chargino mass matrices}
\label{appendix:A}
The fermion sector in our case is different from the usual MSSM structure. The neutralino mass matrix in the basis $(\tilde B, \tilde S, \tilde W^0, \tilde T^0, \tilde H_u^0, \tilde H_d^0)$ looks like
\begin{eqnarray}
M_{\tilde\chi^0} &=&
\begin{pmatrix} 
0 & M_{D_1} & 0 & 0 & \frac{g^{\prime} v_u}{2} & -\frac{g^{\prime} v_d}{2}  \\
M_{D_1} & M_{\Sigma_1} & 0 & 0 & -\frac{\lambda_S v_d}{\sqrt{2}} & -\frac{\lambda_S v_u}{\sqrt{2}} \\
0 & 0 & 0 & M_{D_2} &  -\frac{g v_u}{2} & \frac{g v_d}{2} \\
0 & 0 & M_{D_2} & M_{\Sigma_2} & -\frac{\lambda_T v_d}{2} & -\frac{\lambda_T v_u}{2} \\
\frac{g^{\prime} v_u}{2} & -\frac{\lambda_S v_d}{\sqrt{2}} & -\frac{g v_u}{2}  & -\frac{\lambda_T v_d}{2} & 0 & -\mu \\
-\frac{g^{\prime} v_d}{2} & -\frac{\lambda_S v_u}{\sqrt{2}} & \frac{g v_d}{2}  & -\frac{\lambda_T v_u}{2} & -\mu & 0
\end{pmatrix}.
\end{eqnarray}
Also the chargino mass matrix written in the basis $(\tilde W^-, \tilde T^-, \tilde H_d^-)$ and $(\tilde W^+, \tilde T^+, \tilde H_u^+)$ takes the following shape
\begin{eqnarray}
M_{\tilde\chi^{+}} &=&
\begin{pmatrix}
0 & M_{D_2} & \frac{g v_u}{\sqrt{2}} \\
M_{D_2} & M_{\Sigma_2} & \frac{\lambda_T v_d}{\sqrt{2}} \\
\frac{g v_d}{\sqrt{2}} & -\frac{\lambda_T v_u}{\sqrt{2}} & \mu
\end{pmatrix}.
\end{eqnarray}
\section{Switching on the Yukawa coupling}
\label{Appendix:B}
We now switch on the Yukawa couplings and show how the right handed slepton masses get generated through RGEs. For this the following equations are required to be solved through some approximate means. Such as
\begin{eqnarray}
16\pi^2 \frac{d\mathcal S}{dt} &=& \frac{66}{5}g_1^2 \mathcal S-12 \left|y_t\right|^2\left[\left|\tilde\xi_u\right|^2+\tilde\xi_u^*\mu+\tilde\xi_u\mu^*\right], \label{eq:rgs} \\
16\pi^2 \frac{d\tilde\xi_u}{dt}&\simeq&  3\left|y_t\right|^2 \tilde\xi_u, \\
16\pi^2 \frac{dy_t}{dt} &\simeq& y_t\left[6\left|y_t\right|^2-\frac{16}{3}g_3^2\right]. \label{eq:rgyt}
\end{eqnarray}
Eq.~(\ref{eq:rgyt}) can be simplified to obtain the following form
\begin{eqnarray}
\int_{\Lambda}^{\mu} d\log\left|y_t\right|^2 &=&\frac{3}{4\pi^2}\int_{\Lambda}^{\mu} dt \left|y_t\right|^2-\frac{2 g_3^2}{3\pi^2}\int^{\mu}_{\Lambda} dt.
\end{eqnarray}
This can be further simplified to
\begin{eqnarray}
\frac{3}{4\pi^2} \int_{\Lambda}^{\mu} dt \left|y_t\right|^2 &=& \log\left[\left(\frac{\left|y_t(\mu)\right|^2}{\left|y_t(\Lambda)\right|^2}\right)\left(\frac{\mu}{\Lambda}\right)^{8\alpha_S/3\pi}\right],
\end{eqnarray}
and used for the solution of $\tilde\xi_u$, which we find
\begin{eqnarray}
\tilde\xi_u(\mu) &\simeq& \tilde\xi_u(\Lambda)\left[\left(\frac{y_t(\mu)}{y_t(\Lambda)}\right)\left(\frac{\mu}{\Lambda}\right)^{4\alpha_S/3\pi}\right]^{\frac{1}{2}},
\end{eqnarray}
The final part is the computation of $\mathcal S$ which directly goes into the RGEs of the scalar masses. We simplify by considering
\begin{eqnarray}
16\pi^2 \frac{d\mathcal S}{dt} -\frac{66}{5} g_1^2 \mathcal S \simeq -12\left|y_t\right|^2 \left|\tilde\xi_u\right|^2.
\end{eqnarray}
One can treat the above equation as the most general linear first order ordinary differential equations. The term in the right hand can be regarded as a source or a driving term for the inhomogeneous ordinary differential equation. The solution is straightforward which involves a integrating factor. The final solution can be written in the closed form as
\begin{eqnarray}
\mathcal S(t) &=& -2~\text{exp}\left[2\int^t_{t_0} d\log g_1 (t_1)\right]\left[\int^t d\left|\xi_u(t_3)\right|^2~\text{exp}\left\{-2\int^{t_3} d\log g_1(t_2)\right\}\right],\\
&=& -2 \left[\frac{g_1(t)}{g_1(t_0)}\right]^{2}\left[\int_{t_0}^{t}dt_3 \frac{d\left|\tilde\xi_u(t_3)\right|^2}{dt_3}\left\{g_1(t_3)\right\}^{-2}\right].
\end{eqnarray}
Using integration by parts one can further simply this to obtain
\begin{eqnarray}
\mathcal S(t) &=& -2\left[\frac{g_1(t)}{g_1(t_0)}\right]^2\left[\frac{\left|\tilde\xi_u(t_3)\right|^2}{\left|g_1(t_3)\right|^2}+\frac{b_1}{8\pi^2}\int dt_3\left|\tilde\xi_u(t_3)\right|^2\right]_{t_0}^{t}
\label{eq:solve_S}
\end{eqnarray}
Although, the final result is not really transparent from eq.~(\ref{eq:solve_S}), however it is conspicuous that a non-zero $\mathcal S$ requires a non-zero $\tilde\xi_{u/d}$ at $\MuInt$.
\section{Higgs mass matrix}
\label{Appendix:C}
The scalar mass matrix elements written in the basis $(h_u,h_d)$ after integrating out the adjoint scalars and assuming $\lambda_T=0$, turns out to be
\begin{eqnarray}
M_{11}^2 &=& b_{\mu}\cot\beta+\frac{1}{4}\left[\frac{g^2 M_{\Sigma_2}^2}{4 M_{D_2}^2+M_{\Sigma_2}^2}+\frac{g^{\prime 2} M_{\Sigma_1}^2}{4 M_{D_1}^2+M_{\Sigma_1}^2}\right]v^2 \sin^2\beta+\frac{g^{\prime}\lambda_s}{2\sqrt{2}}\frac{M_{D_1}M_{\Sigma_1}v^2}{4 M_{D_1}^2+M_{\Sigma_1}^2}\left(-2+\cos 2\beta\right)\cot\beta \nonumber \\
&-&\frac{\lambda_s^2}{2}\frac{M_{\Sigma_1}\mu v^2}{4M_{D_1}^2+M_{\Sigma_1}^2}\cos 3\beta\csc\beta+2\sqrt{2}\lambda_s g^{\prime} \frac{M_{D_1}\mu v^2}{4 M_{D_1}^2+M_{\Sigma_1}^2}\sin^2\beta-2\lambda_s^2 \frac{\mu^2 v^2}{4 M_{D_1}^2+M_{\Sigma_1}^2}\sin^2\beta, \nonumber \\
M_{22}^2 &=& b_{\mu}\tan\beta+\frac{1}{4}\left[\frac{g^2 M_{\Sigma_2}^2}{4 M_{D_2}^2+M_{\Sigma_2}^2}+\frac{g^{\prime 2} M_{\Sigma_1}^2}{4 M_{D_1}^2+M_{\Sigma_1}^2}\right]v^2 \cos^2\beta+\frac{g^{\prime}\lambda_s}{2\sqrt{2}}\frac{M_{D_1}M_{\Sigma_1}v^2}{4 M_{D_1}^2+M_{\Sigma_1}^2}\left(2+\cos 2\beta\right)\tan\beta \nonumber \\
&+&\frac{\lambda_s^2}{2}\frac{M_{\Sigma_1}\mu v^2}{4M_{D_1}^2+M_{\Sigma_1}^2}\sin 3\beta\sec\beta+2\sqrt{2}\lambda_s g^{\prime} \frac{M_{D_1}\mu v^2}{4 M_{D_1}^2+M_{\Sigma_1}^2}\cos^2\beta-2\lambda_s^2 \frac{\mu^2 v^2}{4 M_{D_1}^2+M_{\Sigma_1}^2}\cos^2\beta, \nonumber \\
M_{12}^2 &=& -b_{\mu}-\frac{1}{4}\left[\frac{g^2 M_{\Sigma_2}^2}{4 M_{D_2}^2+M_{\Sigma_2}^2}+\frac{g^{\prime 2} M_{\Sigma_1}^2}{4 M_{D_1}^2+M_{\Sigma_1}^2}\right]v^2 \sin\beta \cos\beta+\frac{3 g^{\prime}\lambda_s}{2\sqrt{2}}\frac{M_{D_1}M_{\Sigma_1}v^2}{4 M_{D_1}^2+M_{\Sigma_1}^2}\cos2\beta \nonumber \\
&+&\frac{\lambda_s^2 v^2}{2}\frac{3 M_{\Sigma_1}\mu+\left(4 M_{D_1}^2-2\mu^2\right)\sin 2\beta}{4 M_{D_1}^2+M_{\Sigma_1}^2}.
\end{eqnarray}
Furthermore, integrating out the Dirac adjoint scalars and adding the two minimization equations for $h_u$ and $h_d$ we find
\begin{eqnarray}
\frac{2 b_{\mu}}{\sin 2\beta}&=&\left|\mu_u\right|^2+\left|\mu_d\right|^2+2\lambda_s^2 v^2\frac{M_{D_1}^2-\mu^2}{4 M_{D_1}^2+M_{\Sigma_1}^2}+\frac{g^{\prime}\lambda_s v^2}{\sqrt{2}}\frac{M_{D_1}M_{\Sigma_1}}{4 M_{D_1}^2+M_{\Sigma_1}^2}\cot 2\beta \nonumber \\
&-&\sqrt{2}g^{\prime}\lambda_{s} v^2 \frac{M_{D_1}\mu}{4 M_{D_1}^2+M_{\Sigma_1}^2}\cos 2\beta-\frac{\lambda_{s}^2 v^2}{4} \frac{M_{\Sigma_1}\mu}{4 M_{D_1}^2+M_{\Sigma_1}^2}\left(-7+5\cos 2\beta\right)\cot\beta
\end{eqnarray}
For a fixed right-slepton and higgsino masses,  $\mu_u$ and $\mu_d$ is completely fixed. Therefore, given the values of gaugino masses and $\tan\beta$, one completely fixes $b_\mu$. The size of $b_\mu$ also controls the heavy and charged Higgs masses. Hence, the whole spectrum gets determined. 
\section{RGEs with $\lambda_S$, $\lambda_T$:}
\label{appendix:D}
The inclusion of the superpotential and non-standard soft supersymmetry breaking terms proportional to $\lambda_S$ and $\lambda_T$ modifies the anomalous dimensions of the Higgs fields. This in turn modifies the RGEs of the soft supersymmetry breaking Higgs mass parameters and obviously the $\mu$-term and non-standard supersymmetry breaking terms proportional to $\tilde\xi_u$, $\tilde\xi_d$.
\begin{eqnarray}
\beta\left[\tilde m_{h_u}^2\right]&\rightarrow&\beta\left[\tilde m_{h_u}^2\right]+\left(2\left|\lambda_S\right|^2+6\left|\lambda_T\right|^2\right)\left[\tilde m_{h_u}^2+\tilde m_{h_d}^2+3\left\{|\tilde\xi_d|^2+\tilde\xi_d^*\mu+\tilde\xi_d\mu^*\right\}+\left\{|\tilde\xi_u|^2+\tilde\xi_u^*\mu+\tilde\xi_u\mu^*\right\}\right], \nonumber \\
\beta\left[\tilde m_{h_d}^2\right]&\rightarrow&\beta\left[\tilde m_{h_d}^2\right]+\left(2\left|\lambda_S\right|^2+6\left|\lambda_T\right|^2\right)\left[\tilde m_{h_u}^2+\tilde m_{h_d}^2+3\left\{|\tilde\xi_u|^2+\tilde\xi_u^*\mu+\tilde\xi_u\mu^*\right\}+\left\{|\tilde\xi_d|^2+\tilde\xi_d^*\mu+\tilde\xi_d\mu^*\right\}\right], \nonumber \\
\beta\left[\mu\right] &\rightarrow& \beta\left[\mu\right]+\left(2\left|\lambda_S\right|^2+6\left|\lambda_T\right|^2\right)\mu,\nonumber \\
\beta\left[\tilde\xi_d\right] &\rightarrow& \beta\left[\tilde\xi_d\right]+\left(2\left|\lambda_S\right|^2+6\left|\lambda_T\right|^2\right)\tilde\xi_d,\nonumber \\
\beta\left[\tilde\xi_u\right] &\rightarrow& \beta\left[\tilde\xi_u\right]+\left(2\left|\lambda_S\right|^2+6\left|\lambda_T\right|^2\right)\tilde\xi_u.
\end{eqnarray}
\section{Tadpole issue}
\label{appendix:E}
Another important aspect of our case is the effect of non-standard soft terms in the presence of the singlet. These terms have been traditionally neglected in models with gauge singlets as they could give rise to dangerous tadpole diagrams which might destabilize the hierarchy~\cite{Hetherington:2001bk,Dedes:2000jp}. It is important to see the effect of such terms in GSS.
\begin{figure}[h]
\centering
\includegraphics[width=0.6\textwidth]{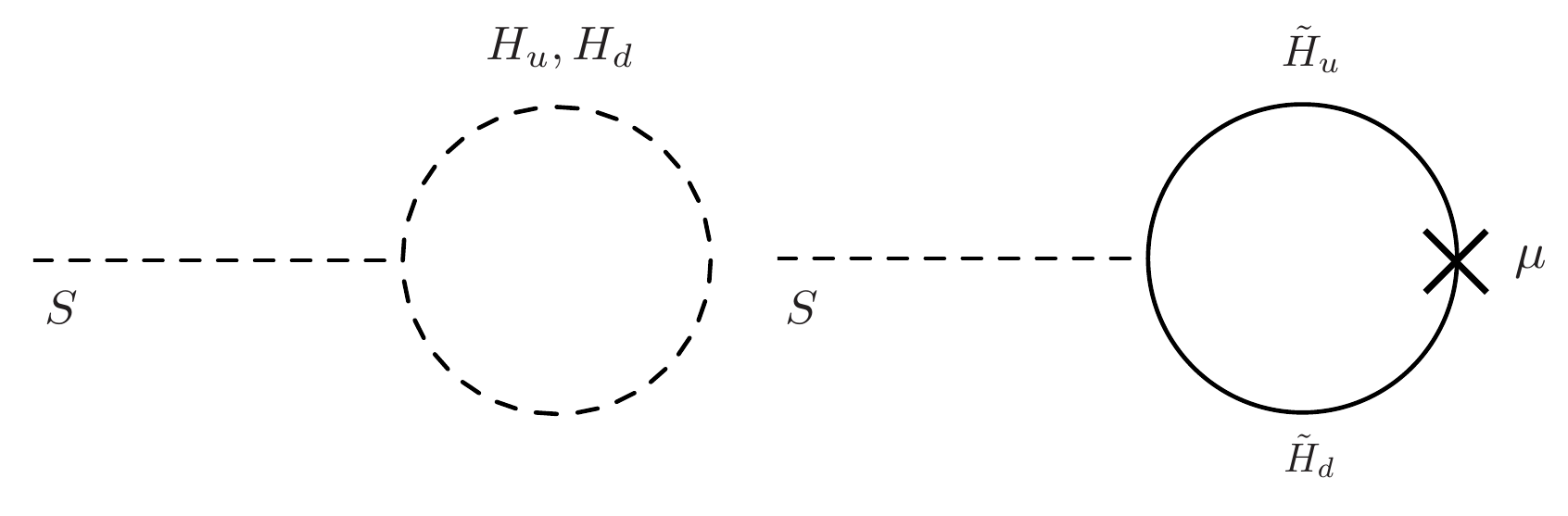}
\caption{Tadpole diagrams originating from the non-standard soft terms in the presence of the gauge singlet chiral superfield.}
\label{fig:sec}
\end{figure}
These diagrams can be evaluated in a straight forward manner and the terms which gives rise to hard breaking are
\begin{equation}
\delta_S = \frac{\lambda_S}{16\pi^2} \MuInt^2 \left[\mu(\MuInt)-\left(\frac{\mu_u^0+\mu_d^0}{2}\right)\right].
\end{equation}
Since we have chosen $\mu(\MuInt)=(\mu_u^0+\mu_d^0)/2$, therefore these tadpole diagrams do not give rise to hard breaking at $\MuInt$.

\bibliography{References.bib}
\bibliographystyle{jhep}
\end{document}